\documentclass{article}
\usepackage{amsmath}
\usepackage{amssymb}

\input{tcilatex}

\begin{document}

\title{Exact Fourier expansion in cylindrical coordinates for the
three-dimensional Helmholtz Green function}
\author{John T. Conway \\
\textit{Department of Engineering and Science, University of Agder,}\\
\textit{Grimstad, Norway} \and Howard S. Cohl \\
\textit{Department of Mathematics, University of Auckland, New Zealand}}
\maketitle
\date{}

\begin{abstract}
A new method is presented for Fourier decomposition of the Helmholtz Green
Function in cylindrical coordinates, which is equivalent to obtaining the
solution of the Helmholtz equation for a general ring source. The Fourier
coefficients of the Helmholtz Green function are split into their half
advanced+half retarded and half advanced$-$half retarded components. Closed
form solutions are given for these components in terms of a Horn function
and a Kamp\'{e} de F\'{e}riet function, respectively. The systems of partial
differential equations associated with these two-dimensional hypergeometric
functions are used to construct a fourth-order ordinary differential
equation which both components satisfy. A second fourth-order ordinary
differential equation for the general Fourier coefficent is derived from an
integral representation of the coefficient, and both differential equations
are shown to be equivalent. Series solutions for the various Fourier
coefficients are also given, mostly in terms of Legendre functions and
Bessel/Hankel functions. These are derived from the closed form
hypergeometric solutions or an integral representation, or both. Numerical
calculations comparing different methods of calculating the Fourier
coefficients are presented.
\end{abstract}

\section{Introduction and overview}

The inhomogeneous Helmholtz wave equation is%
\begin{equation}
\left( \nabla ^{2}+\beta ^{2}\right) \Phi \left( \beta ,\mathbf{r}\right)
=\rho \left( \mathbf{r}\right)  \label{Eqn1}
\end{equation}%
and this has the well known free-space retarded Green function [1, p. 284]%
\begin{equation}
G_{H}\left( \beta ,\mathbf{r-r}^{\prime }\right) =-\frac{\exp \left( i\beta
\mid \mathbf{r}-\mathbf{r}^{\prime }\mid \right) }{4\pi \mid \mathbf{r}-%
\mathbf{r}^{\prime }\mid }  \label{Eqn2}
\end{equation}%
where $\mathbf{r}$ is a field point, $\mathbf{r}^{\prime }$ is a source
point and $\beta $ is the wave number, here considered to be a general
complex number. The free-space Green function (\ref{Eqn2}) is restricted to
values of $\beta $ such that $\mid G_{H}\left( \beta ,\mathbf{r-r}^{\prime
}\right) \mid \rightarrow 0$ as $\mid \mathbf{r-r}^{\prime }\mid \rightarrow
\infty $. For general dispersive waves with $\beta =\alpha +i\sigma $ where $%
\alpha $ and $\sigma $ are real, then $\sigma \geqslant 0$ is a condition
for this to hold. In the limit as $\beta \rightarrow 0$ these equations
reduce to the Poisson equation and its corresponding Green function $%
G_{P}\left( \mathbf{r}-\mathbf{r}^{\prime }\right) $. The general retarded
solution $\Phi \left( \beta ,\mathbf{r}\right) $ of the Helmholtz equation
at a field point $\mathbf{r}$ for a general source density $\rho \left( 
\mathbf{r}\right) $, subject to the boundary condition that $\Phi \left(
\beta ,\mathbf{r}\right) \rightarrow 0$ as $\mid \mathbf{r}\mid \rightarrow
\infty $, is given in terms of the Green function as%
\begin{equation}
\Phi \left( \beta ,\mathbf{r}\right) =\diiint G_{H}\left( \beta ,\mathbf{r}-%
\mathbf{r}^{\prime }\right) \rho \left( \mathbf{r}^{\prime }\right) d^{3}%
\mathbf{r}^{\prime }  \label{Eqn3}
\end{equation}%
where the volume integral is to be taken over all regions of space where the
source density $\rho \left( \mathbf{r}\right) $ is non zero.

Many problems of practical interest have some element of axial symmetry and
are best treated in cylindrical coordinates $\left( r,\phi ,z\right) $, the
Cartesian components $\left( x,y,z\right) $ of $\mathbf{r}$ being related to
the cylindrical components by $\left( x,y,z\right) =\left( r\cos \phi ,r\sin
\phi ,z\right) $. It follows immediately from this relation that the
distance between a source point and a field point is given by%
\begin{equation}
\mid \mathbf{r}-\mathbf{r}^{\prime }\mid =\sqrt{r^{2}+r^{\prime 2}+\left(
z-z^{\prime }\right) ^{2}-2rr^{\prime }\cos \left( \phi -\phi ^{\prime
}\right) }\text{.}  \label{Eqn4}
\end{equation}%
The solution for $\Phi \left( \beta ,\mathbf{r}\right) $ when $\rho \left( 
\mathbf{r}\right) $ is a general circular ring source is of particular
interest, with applications such as circular loop antennas [2], [3], [4],
the acoustics of rotating machinery [5] and acoustic and electromagnetic
scattering [6]. For the simpler Poisson equation most of the analytical
solutions found in the literature for cylindrical geometry are either ring
source solutions or can be easily constructed from them by integration or
summation. Examples are gravitating rings and disks, ring vortices and
vortex disks, and circular current loops and solenoids.

The source density $\rho _{c}\left( \mathbf{r},R,z\right) $ for a thin
circular ring of radius $R$ located in the plane $z=Z$ \ is of the form%
\begin{equation}
\rho _{c}\left( \mathbf{r},R,Z\right) =f\left( \phi \right) \delta \left(
r-R\right) \delta \left( z-Z\right)  \label{Eqn5}
\end{equation}%
where $f\left( \phi \right) $ is the angular distribution of the source
strength around the ring. This can be most conveniently described by a
Fourier series of the form%
\begin{equation}
f\left( \phi \right) =\frac{a_{0}}{2}+\dsum\limits_{m=1}^{\infty }\left(
a_{m}\cos \left( m\phi \right) +b_{m}\sin \left( m\phi \right) \right) \text{%
.}  \label{Eqn6}
\end{equation}%
where the Fourier coefficients $a_{m}$ and $b_{m}$ are given by [7, p. 1066]%
\begin{equation}
a_{m}=\frac{1}{2\pi }\dint\limits_{0}^{2\pi }f\left( \phi \right) \cos
\left( m\phi \right) d\phi  \label{Eqn6a}
\end{equation}%
\begin{equation}
b_{m}=\frac{1}{2\pi }\dint\limits_{0}^{2\pi }f\left( \phi \right) \sin
\left( m\phi \right) d\phi \text{.}  \label{Eqn6b}
\end{equation}%
From equation (\ref{Eqn4}), the Green function (\ref{Eqn3}) is even in the
variable $\psi \equiv \phi ^{\prime }-\phi $, where $\phi $ is the angular
coordinate of $\mathbf{r}$ and $\phi ^{\prime }$ is the angular coordinate
of $\mathbf{r}^{\prime }$. It is convenient to exploit this symmetry when
substituting equations (\ref{Eqn5}) and (\ref{Eqn6}) into (\ref{Eqn3}). From
the identity $f\left( \phi +\psi \right) \equiv f\left( \phi ^{\prime
}\right) $ we obtain 
\begin{equation*}
f\left( \phi ^{\prime }\right) =\frac{a_{0}}{2}+\dsum\limits_{m=1}^{\infty
}\left( a_{m}\cos \left( m\phi \right) +b_{m}\sin \left( m\phi \right)
\right) \cos \left( m\psi \right)
\end{equation*}%
\begin{equation}
+\dsum\limits_{m=1}^{\infty }\left( -a_{m}\sin \left( m\phi \right)
+b_{m}\cos \left( m\phi \right) \right) \sin \left( m\psi \right)
\label{Eqn7}
\end{equation}%
and on substituting equations (\ref{Eqn5}) and (\ref{Eqn7}) into (\ref{Eqn3}%
) and performing the volume integration, the odd terms proportional to $\sin
\left( m\psi \right) $ in equation (\ref{Eqn7}) do not contribute to the
solution $\Phi \left( \beta ,\mathbf{r}\right) $ as $G_{H}\left( \beta ,%
\mathbf{r}-\mathbf{r}^{\prime }\right) $ is even in $\psi $. The remaining
integrals from the even terms can be calculated over the reduced interval
from $0$ to $\pi $. This gives the solution $\Phi _{c}\left( \beta ,\mathbf{%
r,}R,Z\right) $ of the Helmholtz equation for a circular ring source with
general $f\left( \phi \right) $ in the form%
\begin{equation*}
\Phi _{c}\left( \beta ,\mathbf{r},R,Z\right) =-\frac{a_{0}}{2}%
G_{H}^{0}\left( \beta ,r,R,z-Z\right)
\end{equation*}%
\begin{equation}
-\dsum\limits_{m=1}^{\infty }\left( a_{m}\cos \left( m\phi \right)
+b_{m}\sin \left( m\phi \right) \right) G_{H}^{m}\left( \beta ,r,R,z-Z\right)
\label{Eqn8}
\end{equation}%
where%
\begin{equation}
G_{H}^{m}\left( \beta ,r,R,z-Z\right) =\frac{1}{\pi }\dint\limits_{0}^{\pi }%
\frac{\exp \left( i\beta \sqrt{r^{2}+R^{2}+\left( z-Z\right) ^{2}-2rR\cos
\psi }\right) }{\sqrt{r^{2}+R^{2}+\left( z-Z\right) ^{2}-2rR\cos \psi }}\cos
\left( m\psi \right) d\psi  \label{Eqn9}
\end{equation}%
and where the explicit dependence of the solution on the constant ring
parameters $R$ and $Z$ has been introduced in these definitions. Introducing
the Neumann factor $\epsilon _{m}$ such that $\epsilon _{m}=1$ for $m=0$ and 
$\epsilon _{m}=2$ for $m>0$, and defining $b_{0}=0$ allows (\ref{Eqn8}) to
be expressed more concisely as%
\begin{equation}
\Phi _{c}\left( \beta ,\mathbf{r},R,Z\right) =-\frac{1}{2}%
\dsum\limits_{m=0}^{\infty }\left( a_{m}\cos \left( m\phi \right) +b_{m}\sin
\left( m\phi \right) \right) \epsilon _{m}G_{H}^{m}\left( \beta
,r,R,z-Z\right) \text{.}  \label{Eqn9a}
\end{equation}%
Apart from a constant factor, the terms $\epsilon _{m}G_{H}^{m}\left( \beta
,r,R,z-Z\right) $ in (\ref{Eqn9a}) are also the coefficients in the Fourier
expansion of the Green function (\ref{Eqn2}) itself, when the source point
is given by $\mathbf{r}^{\prime }=\left( R,\phi ^{\prime },Z\right) $. From
equations (\ref{Eqn6}), (\ref{Eqn6a}) and (\ref{Eqn6b}) this is given by%
\begin{equation}
G_{H}\left( \beta ,\mathbf{r}-\mathbf{r}^{\prime }\right) =-\frac{1}{4\pi }%
\dsum\limits_{m=0}^{\infty }\epsilon _{m}G_{H}^{m}\left( \beta
,r,R,z-Z\right) \cos \left( m\left( \phi -\phi ^{\prime }\right) \right) 
\text{.}  \label{Eqn10}
\end{equation}%
Thus the solution $\Phi _{c}\left( \beta ,\mathbf{r},R,Z\right) $ of the
Helmholtz equation for a general ring source can be constructed directly
from the coefficients $G_{H}^{m}\left( \beta ,r,R,Z-z\right) $ in the
Fourier expansion of the Green function (\ref{Eqn3}). This provides in large
measure the motivation to analytically construct the Fourier series for the
Helmholtz Green function.

For the Poisson equation with $\beta =0$ the corresponding Fourier expansion
of the Green function has already been given in closed form as [8]:%
\begin{equation}
G_{P}\left( \mathbf{r}-\mathbf{r}^{\prime }\right) =-\frac{1}{4\pi ^{2}\sqrt{%
rR}}\dsum\limits_{m=0}^{\infty }\epsilon _{m}Q_{m-1/2}\left( \omega \right)
\cos \left( m\left( \phi -\phi ^{\prime }\right) \right)  \label{Eqn11}
\end{equation}%
where 
\begin{equation}
\omega =\frac{r^{2}+R^{2}+\left( z-Z\right) ^{2}}{2rR}  \label{Eqn12}
\end{equation}%
is a toroidal variable such that $\omega \geqslant 1$ and the $%
Q_{m-1/2}\left( \omega \right) $ are the Legendre functions of the second
kind and half-integral degree, which are also toroidal harmonics. The
Fourier expansion given by equations (\ref{Eqn11}) and (\ref{Eqn12}) can be
obtained immediately by writing the Green function (\ref{Eqn2}) for $\beta
=0 $ in the form%
\begin{equation}
G_{P}\left( \mathbf{r}-\mathbf{r}^{\prime }\right) =-\frac{1}{4\pi \sqrt{rR}%
\sqrt{2\omega -2\cos \left( \phi -\phi ^{\prime }\right) }}  \label{Eqn13}
\end{equation}%
where $\omega $ is given by (\ref{Eqn12}), and noting that the function $%
Q_{m-1/2}\left( \omega \right) $ has the simple integral representation [7,
eqn 8.713] 
\begin{equation}
Q_{m-1/2}\left( \omega \right) =\dint\limits_{0}^{\pi }\frac{\cos \left(
m\psi \right) d\psi }{\sqrt{2\omega -2\cos \psi }}\text{.}  \label{Eqn14}
\end{equation}%
An alternative derivation of (\ref{Eqn11}) employs the Lipschitz integral
[7, eqn 6.611 1]%
\begin{equation}
\dint\limits_{0}^{\infty }J_{0}\left( sa\right) \exp \left( -s\mid b\mid
\right) ds=\frac{1}{\sqrt{a^{2}+b^{2}}}  \label{Eqn15}
\end{equation}%
and Neumann's addition theorem [9, eqn11.2 1]%
\begin{equation}
J_{0}\left( s\sqrt{r^{2}+R^{2}-2Rr\cos \psi }\right)
=\dsum\limits_{m=0}^{\infty }\epsilon _{m}\cos \left( m\psi \right)
J_{m}\left( sr\right) J_{m}\left( sR\right)  \label{Eqn16}
\end{equation}%
to obtain the well known eigenfunction expansion%
\begin{equation*}
G_{P}\left( \mathbf{r}-\mathbf{r}^{\prime }\right) =-\frac{1}{4\pi }\times
\end{equation*}%
\begin{equation}
\dsum\limits_{m=0}^{\infty }\epsilon _{m}\cos \left( m\left( \phi -\phi
^{\prime }\right) \right) \dint\limits_{0}^{\infty }J_{m}\left( sr\right)
J_{m}\left( sR\right) \exp \left( -s\mid z-Z\mid \right) ds\text{.}
\label{Eqn17}
\end{equation}%
This reduces to (\ref{Eqn11}) on employing the integral [7, eqn 6.612 3],[9,
eqn 13.22]:%
\begin{equation}
\dint\limits_{0}^{\infty }J_{m}\left( sr\right) J_{m}\left( sR\right) \exp
\left( -s\mid Z-z\mid \right) ds=\frac{1}{\pi \sqrt{rR}}Q_{m-1/2}\left(
\omega \right) \text{.}  \label{Eqn18}
\end{equation}%
The generalization of (\ref{Eqn17}) for the Helmholtz case is also well
known [10, p. 888]%
\begin{equation*}
G_{H}\left( \beta ,\mathbf{r}-\mathbf{r}^{\prime }\right) =-\frac{i}{4\pi }%
\times
\end{equation*}%
\begin{equation}
\dsum\limits_{m=0}^{\infty }\epsilon _{m}\cos \left( m\left( \phi -\phi
^{\prime }\right) \right) \dint\limits_{0}^{\infty }\exp \left( i\mid
Z-z\mid \sqrt{\beta ^{2}-s^{2}}\right) J_{m}\left( sr\right) J_{m}\left(
sR\right) \frac{sds}{\sqrt{\beta ^{2}-s^{2}}}\text{.}  \label{Eqn19}
\end{equation}%
This can be similarly obtained from Neumann's theorem by employing the
integral [7, eqn 6.616 2]%
\begin{equation}
\dint\limits_{1}^{\infty }\exp \left( -ax\right) J_{0}\left( b\sqrt{x^{2}-1}%
\right) dx=\frac{\exp \left( -\sqrt{a^{2}+b^{2}}\right) }{\sqrt{a^{2}+b^{2}}}
\label{Eqn20}
\end{equation}%
instead of the Lipschitz integral. Equation (\ref{Eqn19}) gives the Fourier
coefficients of the Helmholtz Green function in the form%
\begin{equation}
G_{H}^{m}\left( \beta ,r,R,z-Z\right) =i\dint\limits_{0}^{\infty }\exp
\left( i\mid Z-z\mid \sqrt{\beta ^{2}-s^{2}}\right) J_{m}\left( sr\right)
J_{m}\left( sR\right) \frac{sds}{\sqrt{\beta ^{2}-s^{2}}}\text{.}
\label{Eqn21}
\end{equation}%
This reduces to (\ref{Eqn17}) in the limit as $\beta \rightarrow 0$ but
unfortunately the integral in (\ref{Eqn21}) is not given in standard tables
for $\beta \neq 0$. Numerical evaluation of this integral requires care, as
the integrand is oscillatory and singular in an infinite range of
integration, though the integrand tends exponentially to zero as $%
s\rightarrow \infty $. Equation (\ref{Eqn9}) is a convenient alternative
numerical evaluation of the Fourier coefficients, provided $m$ is not too
large.

The integrals (\ref{Eqn9}) and (\ref{Eqn21}) contain the additional
parameter $\beta $ which is not contained in (\ref{Eqn14}) and (\ref{Eqn18}%
). As a consequence of this, the closed form generalization of (\ref{Eqn11})
for the Helmholtz case involves two-multidimensional Gaussian hypergeometric
series, and the main purpose of this article is to present these solutions
and various related results. The core idea leading to the solution is
expansion of the exponential in (\ref{Eqn2}) as the absolutely convergent
power series [4]%
\begin{equation}
G_{H}\left( \beta ,\mathbf{r}-\mathbf{r}^{\prime }\right) =-\frac{1}{4\pi }%
\dsum\limits_{n=0}^{\infty }\frac{\left( i\beta \right) ^{n}\mid \mathbf{r}-%
\mathbf{r}^{\prime }\mid ^{n-1}}{n!}  \label{Eqn22}
\end{equation}%
where%
\begin{equation}
\mid \mathbf{r}-\mathbf{r}^{\prime }\mid ^{n-1}=\left( r^{2}+R^{2}+\left(
z-Z\right) ^{2}-2rR\cos \psi \right) ^{\left( n-1\right) /2}\text{.}
\label{Eqn23}
\end{equation}%
Hence%
\begin{equation}
G_{H}^{m}\left( \beta ,r,R,z-Z\right) =\dsum\limits_{n=0}^{\infty }\frac{%
\left( i\beta \right) ^{n}}{n!}I_{m,n}\left( r,R,z-Z\right)  \label{Eqn24}
\end{equation}%
where%
\begin{equation}
I_{m,n}\left( r,R,z-Z\right) =\frac{1}{\pi }\dint\limits_{0}^{\pi }\left(
r^{2}+R^{2}+\left( z-Z\right) ^{2}-2rR\cos \psi \right) ^{\left( n-1\right)
/2}\cos \left( m\psi \right) d\psi \text{.}  \label{Eqn25}
\end{equation}%
The integral (\ref{Eqn25}) can be evaluated as a series by binomial
expansion and this gives a double series for the Fourier coefficient $%
G_{H}^{m}\left( \beta ,r,R,z-Z\right) $. The expansion of (\ref{Eqn25})
gives an infinite number of terms for $n$ even and a finite number of terms
for $n$ odd. These two cases are best treated separately and it is therefore
convenient to split the summation over $n$ in (\ref{Eqn22}) into odd and
even terms. This is equivalent to splitting the Green function (\ref{Eqn2})
such that%
\begin{equation}
G_{H}\left( \beta ,\mathbf{r}-\mathbf{r}^{\prime }\right) =\Lambda
_{+}\left( \beta ,\mathbf{r}-\mathbf{r}^{\prime }\right) +\Lambda _{-}\left(
\beta ,\mathbf{r}-\mathbf{r}^{\prime }\right)  \label{Eqn26}
\end{equation}%
where%
\begin{equation}
\Lambda _{+}\left( \beta ,\mathbf{r}-\mathbf{r}^{\prime }\right) =-\frac{1}{%
8\pi }\left( \frac{\exp \left( i\beta \mid \mathbf{r}-\mathbf{r}^{\prime
}\mid \right) }{\mid \mathbf{r}-\mathbf{r}^{\prime }\mid }+\frac{\exp \left(
-i\beta \mid \mathbf{r}-\mathbf{r}^{\prime }\mid \right) }{\mid \mathbf{r}-%
\mathbf{r}^{\prime }\mid }\right)  \label{Eqn27}
\end{equation}%
is the half advanced+half retarded Green function and%
\begin{equation}
\Lambda _{-}\left( \beta ,\mathbf{r}-\mathbf{r}^{\prime }\right) =-\frac{1}{%
8\pi }\left( \frac{\exp \left( i\beta \mid \mathbf{r}-\mathbf{r}^{\prime
}\mid \right) }{\mid \mathbf{r}-\mathbf{r}^{\prime }\mid }-\frac{\exp \left(
-i\beta \mid \mathbf{r}-\mathbf{r}^{\prime }\mid \right) }{\mid \mathbf{r}-%
\mathbf{r}^{\prime }\mid }\right)  \label{Eqn28}
\end{equation}%
is the half advanced$-$half retarded Green function. The corresponding
Fourier coefficients are split in the same manner such that%
\begin{equation}
\Lambda _{+}^{m}\left( \beta ,r,R,z-Z\right) =\frac{1}{2}\left(
G_{H}^{m}\left( \beta ,r,R,z-Z\right) +G_{H}^{m}\left( -\beta
,r,R,z-Z\right) \right)  \label{Eqn29}
\end{equation}%
\begin{equation}
\Lambda _{-}^{m}\left( \beta ,r,R,z-Z\right) =\frac{1}{2}\left(
G_{H}^{m}\left( \beta ,r,R,z-Z\right) +G_{H}^{m}\left( -\beta
,r,R,z-Z\right) \right) .  \label{Eqn30}
\end{equation}%
For real $\beta $, splitting the Green function in this way is equivalent to
dividing it into its real and imaginary parts, but this is not the case for
general complex $\beta $. It is shown in Section 2 that the Fourier
coefficients in (\ref{Eqn29})\ and (\ref{Eqn30}) are given respectively by%
\begin{equation*}
\bar{\Lambda}_{+}^{m}\left( rR,\gamma ,k\right) =
\end{equation*}%
\begin{equation}
\frac{\Gamma \left( m+1/2\right) }{m!\sqrt{\pi rR}}\left( \frac{k}{2}\right)
^{2m+1}\mathrm{H}_{3}\left( m+1/2,m+1/2,2m+1,k^{2},\frac{\gamma ^{2}}{4}%
\right)  \label{Eqn31}
\end{equation}%
and%
\begin{equation*}
\bar{\Lambda}_{-}^{m}\left( rR,\gamma ,k\right) =\frac{i}{\sqrt{rR}\left(
2m+1\right) !}\left( \frac{\gamma k}{2}\right) ^{2m+1}\times
\end{equation*}%
\begin{equation}
F_{1:1;0}^{0:1;0}\left[ 
\begin{array}{c}
-: \\ 
m+3/2:%
\end{array}%
\begin{array}{c}
m+1/2; \\ 
2m+1;%
\end{array}%
\begin{array}{c}
-; \\ 
-;%
\end{array}%
\frac{\gamma ^{2}k^{2}}{4},\frac{-\gamma ^{2}}{4}\right]  \label{Eqn32}
\end{equation}%
where%
\begin{equation}
k=\sqrt{\frac{4rR}{\left( r+R\right) ^{2}+\left( z-Z\right) ^{2}}}
\label{Eqn33}
\end{equation}%
\begin{equation}
\gamma =\beta \sqrt{\left( r+R\right) ^{2}+\left( z-Z\right) ^{2}}
\label{Eqn34}
\end{equation}%
and 
\begin{equation}
\bar{\Lambda}_{\pm }^{m}\left( rR,\gamma ,k\right) \equiv \Lambda _{\pm
}^{m}\left( \beta ,r,R,z-Z\right) \text{.}  \label{Eqn35}
\end{equation}%
The variable $k$ is the usual modulus contained in elliptic integral
solutions of elementary ring problems and is related to the toroidal
variable $\omega $ by%
\begin{equation}
\omega =\frac{2-k^{2}}{k^{2}}\text{.}  \label{Eqn36}
\end{equation}%
The function $\mathrm{H}_{3}$ in equation (\ref{Eqn31}) is one of the
standard Horn functions [11, eqn 5.7.1 31] and is equivalent to the double
hypergeometric series%
\begin{equation*}
\mathrm{H}_{3}\left( m+1/2,m+1/2,2m+1,k^{2},\frac{\gamma ^{2}}{4}\right) =
\end{equation*}%
\begin{equation}
\dsum\limits_{n=0}^{\infty }\dsum\limits_{p=0}^{\infty }\frac{\left(
m+1/2\right) _{n-p}\left( m+1/2\right) _{n}}{\left( 2m+1\right) _{n}n!p!}%
\left( k^{2}\right) ^{n}\left( \frac{\gamma ^{2}}{4}\right) ^{p}\text{.}
\label{Eqn37}
\end{equation}%
The Kamp\'{e} de F\'{e}riet function [12, p. 27] in (\ref{Eqn32}) is
equivalent to the double hypergeometric series%
\begin{equation*}
F_{1:1;0}^{0:1;0}\left[ 
\begin{array}{c}
-: \\ 
m+3/2:%
\end{array}%
\begin{array}{c}
m+1/2; \\ 
2m+1;%
\end{array}%
\begin{array}{c}
-; \\ 
-;%
\end{array}%
\frac{\gamma ^{2}k^{2}}{4},\frac{-\gamma ^{2}}{4}\right] =
\end{equation*}%
\begin{equation}
\dsum\limits_{n=0}^{\infty }\dsum\limits_{p=0}^{\infty }\frac{\left(
m+1/2\right) _{n}}{\left( m+3/2\right) _{n+p}\left( 2m+1\right) _{n}n!p!}%
\left( \frac{\gamma ^{2}k^{2}}{4}\right) ^{n}\left( \frac{-\gamma ^{2}}{4}%
\right) ^{p}\text{.}  \label{Eqn38}
\end{equation}

The integral (\ref{Eqn25}) can also be evaluated using an integral
representation for the associated Legendre function of the first kind, and
it is shown in Appendix A that this gives the series expansion: 
\begin{equation*}
\hat{G}_{H}^{m}\left( rR,\lambda ,\omega \right) =\frac{\left( -1\right) ^{m}%
}{\left( \omega ^{2}-1\right) ^{1/4}\sqrt{2rR}}\times
\end{equation*}%
\begin{equation}
\dsum\limits_{n=0}^{\infty }\frac{\left( i\lambda \left( \omega
^{2}-1\right) ^{1/4}\right) ^{n}}{n!}\frac{\Gamma \left( \left( n+1\right)
/2\right) }{\Gamma \left( m+\left( n+1\right) /2\right) }P_{\left(
n-1\right) /2}^{m}\left( \frac{\omega }{\sqrt{\omega ^{2}-1}}\right)
\label{Eqn39}
\end{equation}%
where 
\begin{equation}
\lambda =\beta \sqrt{2rR}  \label{Eqn40}
\end{equation}%
and%
\begin{equation}
\hat{G}_{H}^{m}\left( rR,\lambda ,\omega \right) \equiv G_{H}^{m}\left(
\beta ,r,R,z-Z\right) \text{.}  \label{Eqn41}
\end{equation}%
The Legendre function in equation (\ref{Eqn39}) reduces to an associated
Legendre polynomial for odd $n$. The series in (\ref{Eqn39}) can be split
into even and odd terms such that%
\begin{equation}
\hat{G}_{H}^{m}\left( rR,\lambda ,\omega \right) =\frac{1}{2}\left( \hat{%
\Lambda}_{+}^{m}\left( rR,\lambda ,k\right) +\hat{\Lambda}_{-}^{m}\left(
rR,\lambda ,\omega \right) \right)  \label{Eqn42}
\end{equation}%
where 
\begin{equation}
\hat{\Lambda}_{\pm }^{m}\left( rR,\lambda ,\omega \right) \equiv \Lambda
_{\pm }^{m}\left( \beta ,r,R,z-Z\right)  \label{Eqn43}
\end{equation}%
and it is shown in Appendix A that the even and odd series can be expressed
respectively as:%
\begin{equation}
\hat{\Lambda}_{+}^{m}\left( rR,\lambda ,\omega \right) =\frac{\left(
-1\right) ^{m}}{\sqrt{rR}}\dsum\limits_{p=0}^{\infty }\left( \frac{\lambda
^{2}\sqrt{\omega ^{2}-1}}{4}\right) ^{p}\frac{Q_{m-1/2}^{p}\left( \omega
\right) }{p!\Gamma \left( p-m+1/2\right) \Gamma \left( p+m+1/2\right) }
\label{Eqn44}
\end{equation}%
\begin{equation}
\hat{\Lambda}_{-}^{m}\left( rR,\lambda ,\omega \right) =\frac{\left(
-1\right) ^{m}}{\sqrt{rR}}\dsum\limits_{p=0}^{\infty }\left( \frac{\lambda
^{2}\sqrt{\omega ^{2}-1}}{4}\right) ^{p+m+1/2}\frac{Q_{m-1/2}^{p+m+1/2}%
\left( \omega \right) }{p!\Gamma \left( p+m+3/2\right) \Gamma \left(
p+2m+1\right) }\text{.}  \label{Eqn45}
\end{equation}%
In equation (\ref{Eqn45}) the Legendre function is purely imaginary for real 
$\lambda $. In the static limit as $\lambda \rightarrow 0$ then $\hat{\Lambda%
}_{-}^{m}\left( rR,0,\omega \right) =0$ and from the gamma function identity
[7, eqn 8.334 2] 
\begin{equation}
\Gamma \left( 1/2-m\right) \Gamma \left( 1/2+m\right) =\left( -1\right)
^{m}\pi \text{ }\left[ m\in 
\mathbb{N}
_{0}\right]  \label{Eqn46}
\end{equation}%
then equation (\ref{Eqn44}) reduces to%
\begin{equation}
\hat{\Lambda}_{+}^{m}\left( rR,0,\omega \right) =\frac{Q_{m-1/2}\left(
\omega \right) }{\pi \sqrt{rR}}\text{ }\left[ m\in 
\mathbb{N}
_{0}\right]  \label{Eqn47}
\end{equation}%
as it must do for consistency with (\ref{Eqn11}).

The solutions in terms of two-dimensional hypergeometric functions defined
by Equations (\ref{Eqn31})-(\ref{Eqn35}) and (\ref{Eqn37})-(\ref{Eqn38}) can
be summed over either index to give the solutions as series of special
functions. It is shown in Section 3 that summation over the index $n$ in
equation (\ref{Eqn37}) gives equation (\ref{Eqn44}), exactly as given by the
integral representation.\ However, summation over the index $n$ in equation (%
\ref{Eqn38}) gives instead the series solution%
\begin{equation*}
\bar{\Lambda}_{-}^{m}\left( rR,\gamma ,k\right) =\frac{i\sqrt{\pi }}{m!\sqrt{%
rR}}\left( \frac{\gamma k}{4}\right) ^{2m+1}\times
\end{equation*}%
\begin{equation}
\dsum\limits_{p=0}^{\infty }\frac{1}{\Gamma \left( p+m+3/2\right) p!}\left( 
\frac{-\gamma ^{2}}{4}\right) ^{p}{}{}_{1}F_{2}\left( m+1/2;n+m+3/2,2m+1;%
\frac{\gamma ^{2}k^{2}}{4}\right) \text{.}  \label{Eqn48}
\end{equation}%
A hypergeometric identity to reduce the hypergeometric function in equation (%
\ref{Eqn48}) to other well-known special functions does not seem to be
available in standard tabulations. It might nevertheless be conjectured that
(\ref{Eqn48}) could somehow be reducible to equation (\ref{Eqn45}), but this
is not in fact the case. It is easily verified numerically that although
equations (\ref{Eqn45}) and (\ref{Eqn48}) both converge rapidly to the same
limit, the individual terms do not match. Hence, equation (\ref{Eqn48}) is a
distinct series from equation (\ref{Eqn45}). It is also shown in Section 3
that summation over the index $p$ in equations (\ref{Eqn37}) and (\ref{Eqn38}%
) gives the Bessel function series: 
\begin{equation}
\bar{\Lambda}_{+}^{m}\left( rR,\gamma ,k\right) =\frac{-1}{2\sqrt{rR}}%
\dsum\limits_{n=0}^{\infty }\frac{\Gamma \left( n+m+1/2\right) }{\Gamma
\left( n+2m+1\right) n!}\left( \frac{\gamma k^{2}}{2}\right)
^{n+m+1/2}Y_{n+m+\frac{1}{2}}\left( \gamma \right)  \label{Eqn49}
\end{equation}%
\begin{equation}
\bar{\Lambda}_{-}^{m}\left( rR,\gamma ,k\right) =\frac{i}{2\sqrt{rR}}%
\sum_{n=0}^{\infty }\frac{\Gamma \left( n+m+1/2\right) }{\Gamma \left(
n+2m+1\right) n!}\left( \frac{\gamma k^{2}}{2}\right) ^{n+m+1/2}J_{n+m+\frac{%
1}{2}}\left( \gamma \right)  \label{Eqn50}
\end{equation}%
and these two series can be conveniently combined to give a series of Hankel
functions of the first kind:%
\begin{equation}
\bar{G}_{H}^{m}\left( rR,\gamma ,k\right) =\frac{i}{2\sqrt{rR}}%
\dsum\limits_{n=0}^{\infty }\frac{\Gamma \left( n+m+1/2\right) }{\Gamma
\left( n+2m+1\right) n!}\left( \frac{\gamma k^{2}}{2}\right)
^{n+m+1/2}H_{n+m+\frac{1}{2}}^{\left( 1\right) }\left( \gamma \right) \text{.%
}  \label{Eqn51}
\end{equation}

From the solutions (\ref{Eqn31}) and (\ref{Eqn32}) it can be seen that
dimensionless Fourier coefficients defined by%
\begin{equation}
g_{\pm }^{m}\left( k^{2},\gamma ^{2}/4\right) \equiv \sqrt{rR}\bar{\Lambda}%
_{\pm }^{m}\left( rR,\gamma ,k\right)  \label{Eqn52}
\end{equation}%
depend only on the two dimensionless variables $x\equiv k^{2}$ and $y\equiv
\gamma ^{2}/4$. The functions $g_{\pm }^{m}\left( x,y\right) $ are given
explicitly by equations (\ref{Eqn31}) and (\ref{Eqn32}) as:%
\begin{equation}
g_{+}^{m}\left( x,y\right) =\frac{\Gamma \left( m+1/2\right) }{2^{2m+1}m!%
\sqrt{\pi }}x^{\alpha }\mathrm{H}_{3}\left( \alpha ,\alpha ,2\alpha
,x,y\right)  \label{Eqn53}
\end{equation}%
and%
\begin{equation}
g_{-}^{m}\left( x,y\right) =\frac{i}{\left( 2m+1\right) !}\left( xy\right)
^{\alpha }F_{1:1;0}^{0:1;0}\left[ 
\begin{array}{c}
-: \\ 
\alpha +1:%
\end{array}%
\begin{array}{c}
\alpha ; \\ 
2\alpha ;%
\end{array}%
\begin{array}{c}
-; \\ 
-;%
\end{array}%
xy,-y\right]  \label{Eqn54}
\end{equation}%
where%
\begin{equation}
\alpha =m+1/2\text{.}  \label{Eqn55}
\end{equation}%
Two-dimensional hypergeometric series such as (\ref{Eqn53}) and (\ref{Eqn54}%
) are associated with pairs of partial differential equations [11, section
5.9] and these can be used to construct ordinary differential equations for $%
g_{\pm }^{m}\left( x,y\right) $ with $y$ fixed and $x$ as the independent
variable. It is shown in Section 4 that for constant $y$ the coefficients $%
g_{\pm }^{m}\left( x,y\right) $ both satisfy the same fourth-order ordinary
differential equation in $x$:%
\begin{equation*}
\left( 1-x\right) x^{4}\frac{d^{4}g_{\pm }^{m}}{dx^{4}}+\left( 6-9x\right)
x^{3}\frac{d^{3}g_{\pm }^{m}}{dx^{3}}+\left( \alpha \left( 1-\alpha \right)
+6-18x-xy+2y\right) x^{2}\frac{d^{2}g_{\pm }^{m}}{dx^{2}}+
\end{equation*}%
\begin{equation}
\left( 2\alpha \left( 1-\alpha \right) -2x\left( 3-y\right) \right) x\frac{%
dg_{\pm }^{n}}{dx}+\left( y^{2}+2\alpha \left( 1-\alpha \right) y-3\alpha
\left( \alpha +1\right) \left( \alpha +2\right) \right) g_{\pm }^{m}=0\text{.%
}  \label{Eqn56}
\end{equation}

In Section 5 an integral representation is derived for 
\begin{equation}
\hat{y}_{m}\left( \lambda ,\omega \right) \equiv \pi \sqrt{2rR}\hat{G}%
_{H}^{m}\left( rR,\lambda ,\omega \right)  \label{Eqn57}
\end{equation}%
and this is used to derive a fourth-order ordinary differential equation for 
$\hat{y}_{m}\left( \lambda ,\omega \right) $ in terms of $\omega $: 
\begin{equation}
(1-\omega ^{2})\frac{d^{4}\hat{y}_{m}}{d\omega ^{4}}-6\omega \frac{d^{3}\hat{%
y}_{m}}{d\omega ^{3}}+(m^{2}-\frac{\lambda ^{2}\omega }{2}-\frac{25}{4})%
\frac{d^{2}\hat{y}_{m}}{d\omega ^{2}}-\lambda ^{2}\frac{d\hat{y}_{m}}{%
d\omega }-\left( \frac{\lambda ^{2}}{4}\right) ^{2}\hat{y}_{m}=0\text{.}
\label{Eqn58}
\end{equation}%
In the static limit as $\lambda \rightarrow 0$, equation (\ref{Eqn58})
reduces to:%
\begin{equation}
\frac{d^{2}}{d\omega ^{2}}\left[ \left( 1-\omega ^{2}\right) \frac{d^{2}\hat{%
y}_{m}}{d\omega ^{2}}-2\omega \frac{d\hat{y}_{m}}{d\omega }+\left( m^{2}-%
\frac{1}{4}\right) \hat{y}_{m}\right] =0  \label{Eqn59}
\end{equation}%
where%
\begin{equation}
\left( 1-\omega ^{2}\right) \frac{d^{2}\hat{y}_{m}}{d\omega ^{2}}-2\omega 
\frac{d\hat{y}_{m}}{d\omega }+\left( m^{2}-\frac{1}{4}\right) \hat{y}_{m}=0
\label{Eqn60}
\end{equation}%
is Legendre's equation of degree $m-1/2$ [7, eqn 8.820]. It is also shown in
Section 5 that the differential equations (\ref{Eqn56}) and (\ref{Eqn58}),
obtained by quite different routes, are equivalent. The special functions
used in the analysis are given in Table 1. 
\begin{table}[tbp]
$%
\begin{array}{cc}
\text{Symbol} & \text{Special Function} \\ 
\left( a\right) _{n} & \text{Pochammer symbol} \\ 
B\left( x,y\right) & \text{Beta function} \\ 
{}{}_{2}F_{1}\left( a,b;c;x\right) & \text{Gauss hypergeometric function} \\ 
{}{}_{1}F_{2}(a;b,c;x) & \text{A hypergeometric function} \\ 
F_{1:1;0}^{0:1;0}\left[ 
\begin{array}{c}
-: \\ 
b:%
\end{array}%
\begin{array}{c}
a; \\ 
c;%
\end{array}%
\begin{array}{c}
-; \\ 
-;%
\end{array}%
x,y\right] & \text{A Kamp\'{e} de F\'{e}riet function} \\ 
H_{\nu }^{\left( 1\right) }\left( x\right) & \text{Hankel function of the
first kind} \\ 
\text{H}_{3}\left( a,b,c,x,y\right) & \text{The H}_{3}\text{ confluent Horn
function} \\ 
J_{\nu }(x) & \text{Bessel function of the first kind} \\ 
P_{\nu }^{\mu }(x) & \text{Associated Legendre function of the first kind}
\\ 
Q_{\nu }^{\mu }(x) & \text{Associated Legendre function of the second kind}
\\ 
Y_{\nu }\left( x\right) & \text{Bessel function of the second kind} \\ 
\Gamma (x) & \text{Gamma function} \\ 
\delta \left( x\right) & \text{Dirac delta function}%
\end{array}%
$%
\caption{Special Functions Used}
\end{table}

Recurrence relations for the Fourier coefficients for the Helmholtz equation
were investigated by Matviyenko [6], but the closed form solutions and
differential equations presented here appear to be new. Werner [3] presented
an expansion of the Fourier coefficient as a series of spherical Hankel
functions, superficially similar to equation (\ref{Eqn51}), but the two
expansions are distinct. The two-dimensional hypergeometric series approach
applied here to obtain the Fourier expansion for the Helmholtz Green
function has recently been applied to obtain the Fourier expansion in terms
of the amplitude $\phi $ for the Legendre incomplete elliptic integral of
the third kind [13].

The numerical performance of the various expressions for the Fourier
coefficients was investigated using Mathematica$^{\registered }$ [14] and
this is examined in Appendix C.

\section{Solution in terms of two-dimensional hypergeometric series}

The power series expansion (\ref{Eqn24}) for the Fourier coefficient $%
G_{H}^{m}\left( \beta ,r,R,z-Z\right) $ can be expressed in the form%
\begin{equation}
\bar{G}_{H}^{m}\left( rR,\gamma ,k\right) =\frac{\left( -1\right) ^{m}k}{\pi 
\sqrt{rR}}\dsum\limits_{n=0}^{\infty }\frac{\left( i\gamma \right) ^{n}}{n!}%
\dint\limits_{0}^{\pi /2}\cos \left( 2m\theta \right) \left( 1-k^{2}\sin
^{2}\theta \right) ^{\left( n-1\right) /2}d\theta  \label{EqnA1}
\end{equation}%
where%
\begin{equation}
G_{H}^{m}\left( \beta ,r,R,z-Z\right) =\bar{G}_{H}^{m}\left( rR,\gamma
\left( \beta ,r,R,z-Z\right) ,k\left( r,R,z-Z\right) \right)  \label{EqnA2}
\end{equation}%
and $k$ and $\gamma $ are defined by (\ref{Eqn33})\ and (\ref{Eqn34}). The
term $\left( 1-k^{2}\sin ^{2}\theta \right) ^{\left( n-1\right) /2}$ in (\ref%
{EqnA1}) can be expanded binomially to give $\bar{G}_{H}^{m}\left( rR,\gamma
,k\right) $ as a double series containing integrals of the form%
\begin{equation}
\bar{I}_{m,p}=\dint\limits_{0}^{\pi /2}\sin ^{2p}\theta \cos \left( 2m\theta
\right) d\theta \text{.}  \label{EqnA3}
\end{equation}%
This integral is given by Gradshteyn and Ryzhik [7, eqns 3.631 8,12] in a
form which can be recast as%
\begin{equation}
I_{m,p}=\frac{\left( -1\right) ^{m}\pi }{2^{2p+1}\left( 2p+1\right) }\frac{1%
}{B\left( p+m+1,p-m+1\right) }\text{ for }p\geqslant m  \label{EqnA4}
\end{equation}%
\begin{equation}
I_{m,p}=0\text{ for }p<m  \label{EqnA5}
\end{equation}%
and expressing the beta function in (\ref{EqnA4}) in terms of gamma
functions and employing the duplication theorem [7, eqn 8.335 1]%
\begin{equation}
\Gamma \left( 2x\right) =\frac{2^{2x-1}}{\sqrt{\pi }}\Gamma \left( x\right)
\Gamma \left( x+1/2\right)  \label{EqnA6}
\end{equation}%
gives after some reduction the alternative form%
\begin{equation}
I_{m,p}=\frac{\left( -1\right) ^{m}\sqrt{\pi }}{2}\frac{\Gamma \left(
p+1/2\right) p!}{\Gamma \left( p+m+1\right) \Gamma \left( p-m+1\right) }%
\text{ for }p\geqslant m\text{.}  \label{EqnA7}
\end{equation}%
The binomial expansion of (\ref{EqnA1}) gives an infinite series for $n$
zero or even and a finite sum for $n$ odd, and these two cases must be
treated separately. It is therefore convenient to split the series for $\bar{%
G}_{H}^{m}$ such that 
\begin{equation}
\bar{G}_{H}^{m}\left( rR,\gamma ,k\right) =\bar{\Lambda}_{+}^{m}\left(
rR,\gamma ,k\right) +\bar{\Lambda}_{-}^{m}\left( rR,\gamma ,k\right)
\label{EqnA8}
\end{equation}%
and on employing equation (\ref{EqnA6}) the divided series are given by%
\begin{equation}
\bar{\Lambda}_{+}^{m}\left( rR,\gamma ,k\right) =\frac{\left( -1\right) ^{m}k%
}{\sqrt{\pi rR}}\dsum\limits_{n=0}^{\infty }\frac{\left( -\gamma
^{2}/4\right) ^{n}}{\Gamma \left( n+1/2\right) n!}\dint\limits_{0}^{\pi
/2}\cos \left( 2m\theta \right) \left( 1-k^{2}\sin ^{2}\theta \right)
^{n-1/2}d\theta  \label{EqnA9}
\end{equation}%
\begin{equation}
\bar{\Lambda}_{-}^{m}\left( rR,\gamma ,k\right) =\frac{i\left( -1\right)
^{m}k\gamma }{2\sqrt{\pi rR}}\dsum\limits_{n=m}^{\infty }\frac{\left(
-\gamma ^{2}/4\right) ^{n}}{\Gamma \left( n+3/2\right) n!}%
\dint\limits_{0}^{\pi /2}\cos \left( 2m\theta \right) \left( 1-k^{2}\sin
^{2}\theta \right) ^{n}d\theta \text{.}  \label{EqnA10}
\end{equation}%
Binomial expansion of the integrals in equations (\ref{EqnA9})\ and (\ref%
{EqnA10}) gives respectively%
\begin{equation}
\bar{\Lambda}_{+}^{m}\left( rR,\gamma ,k\right) =\frac{\left( -1\right) ^{m}k%
}{\sqrt{\pi rR}}\dsum\limits_{n=0}^{\infty }\dsum\limits_{s=m}^{\infty }%
\frac{\left( -\gamma ^{2}/4\right) ^{n}}{\Gamma \left( n+1/2\right) n!}\frac{%
\Gamma \left( s-n+1/2\right) }{\Gamma \left( 1/2-n\right) }\frac{\left(
k^{2}\right) ^{s}}{s!}\bar{I}_{m,s}  \label{EqnA11}
\end{equation}%
\begin{equation}
\bar{\Lambda}_{-}^{m}\left( rR,\gamma ,k\right) =\frac{i\left( -1\right)
^{m}k\gamma }{2\sqrt{\pi rR}}\dsum\limits_{n=0}^{\infty
}\dsum\limits_{s=m}^{n}\frac{\left( -\gamma ^{2}/4\right) ^{n}}{\Gamma
\left( n+3/2\right) }\frac{\left( -k^{2}\right) ^{s}}{s!\left( n-s\right) !}%
I_{m,s}  \label{EqnA12}
\end{equation}%
and employing the explicit formula (\ref{EqnA7})\ for $I_{m,s}$ in (\ref%
{EqnA11}) and (\ref{EqnA12}) gives respectively%
\begin{equation*}
\bar{\Lambda}_{+}^{m}\left( rR,\gamma ,k\right) =\frac{k}{2\pi \sqrt{rR}}%
\times
\end{equation*}%
\begin{equation}
\dsum\limits_{n=0}^{\infty }\dsum\limits_{s=m}^{\infty }\frac{\Gamma \left(
s-n+1/2\right) \Gamma \left( s+1/2\right) }{\Gamma \left( s+m+1\right)
\Gamma \left( s-m+1\right) }\frac{\left( \gamma ^{2}/4\right) ^{n}\left(
k^{2}\right) ^{s}}{n!}  \label{EqnA13}
\end{equation}%
\begin{equation*}
\bar{\Lambda}_{-}^{m}\left( rR,\gamma ,k\right) =\frac{ik\gamma }{4\sqrt{rR}}%
\times
\end{equation*}%
\begin{equation}
\dsum\limits_{n=m}^{\infty }\dsum\limits_{s=m}^{n}\frac{\Gamma \left(
s+1/2\right) \left( -\gamma ^{2}/4\right) ^{n}\left( -k\right) ^{s}}{\left(
n-s\right) !\Gamma \left( n+3/2\right) \Gamma \left( s+m+1\right) \Gamma
\left( s-m+1\right) }  \label{EqnA14}
\end{equation}%
where the gamma identity (\ref{Eqn46}) has been used to simplify equation (%
\ref{EqnA13}).

\subsection{The series for $\bar{\Lambda}_{+}^{m}\left( rR,\protect\gamma %
,k\right) $}

The substitution $s=p+m$ in equation (\ref{EqnA13}) yields after some
reduction the double series:%
\begin{equation*}
\bar{\Lambda}_{+}^{m}\left( rR,\gamma ,k\right) =\frac{\Gamma \left(
m+1/2\right) }{m!\sqrt{\pi rR}}\left( \frac{k}{2}\right) ^{2m+1}\times
\end{equation*}%
\begin{equation}
\dsum\limits_{n=0}^{\infty }\dsum\limits_{p=0}^{\infty }\frac{\left(
m+1/2\right) _{p-n}\left( m+1/2\right) _{p}}{\left( 2m+1\right) _{p}}\frac{%
\left( k^{2}\right) ^{p}\left( \gamma ^{2}/4\right) ^{n}}{n!p!}
\label{EqnA15}
\end{equation}%
where in (\ref{EqnA15})%
\begin{equation}
\left( a\right) _{p}\equiv \frac{\Gamma \left( a+p\right) }{\Gamma \left(
a\right) }  \label{EqnA16}
\end{equation}%
is the Pochhammer symbol. The double hypergeometric function in (\ref{EqnA15}%
) can be identified as one of the confluent Horn functions [11, eqn 5.7.1
31] and hence%
\begin{equation*}
\bar{\Lambda}_{+}^{m}\left( rR,\gamma ,k\right) =\frac{\Gamma \left(
m+1/2\right) }{m!\sqrt{\pi rR}}\left( \frac{k}{2}\right) ^{2m+1}\times
\end{equation*}%
\begin{equation}
\mathrm{H}_{3}\left( m+1/2,m+1/2,2m+1,k^{2},\frac{\gamma ^{2}}{4}\right) 
\text{.}  \label{EqnA17}
\end{equation}%
The convergence condition given in [11, eqn 5.7.1 31] for the double series
in equation (\ref{EqnA15}) is $k^{2}<1$, which always holds. The order of
summation in equation (\ref{EqnA15}) can be reversed, but the order of the
arguments $k^{2}$ and $\gamma ^{2}/4$ in equation (\ref{EqnA17}) cannot be
exchanged.

\subsection{The series for $\bar{\Lambda}_{-}^{m}\left( rR,\protect\gamma %
,k\right) $}

$\allowbreak $Equation (\ref{EqnA14}) can be converted to a doubly infinite
series by reversing the order of summation, which gives%
\begin{equation*}
\bar{\Lambda}_{-}^{m}\left( rR,\gamma ,k\right) =\frac{ik\gamma }{4\sqrt{rR}}%
\times
\end{equation*}%
\begin{equation}
\dsum\limits_{s=m}^{\infty }\dsum\limits_{n=s}^{\infty }\frac{\Gamma \left(
s+1/2\right) \left( -\gamma ^{2}/4\right) ^{n}\left( -k^{2}\right) ^{s}}{%
\left( n-s\right) !\Gamma \left( n+3/2\right) \Gamma \left( s+m+1\right)
\Gamma \left( s-m+1\right) }\text{.}  \label{EqnA18}
\end{equation}%
The substitution $n=s+p$ in (\ref{EqnA18}) gives%
\begin{equation*}
\bar{\Lambda}_{-}^{m}\left( rR,\gamma ,k\right) =\frac{ik\gamma }{4\sqrt{rR}}%
\times
\end{equation*}%
\begin{equation}
\dsum\limits_{s=m}^{\infty }\dsum\limits_{p=0}^{\infty }\frac{\Gamma \left(
s+1/2\right) }{p!\Gamma \left( p+s+3/2\right) \Gamma \left( s+m+1\right)
\Gamma \left( s-m+1\right) }\left( \frac{\gamma ^{2}k^{2}}{4}\right)
^{s}\left( -\frac{\gamma ^{2}}{4}\right) ^{p}  \label{EqnA19}
\end{equation}%
and the further substitution $s=n+m$ in (\ref{EqnA19}) gives%
\begin{equation*}
\bar{\Lambda}_{-}^{m}\left( rR,\gamma ,k\right) =\frac{i}{2\sqrt{rR}}\times
\left( \frac{\gamma ^{2}k^{2}}{4}\right) ^{m+1/2}\times
\end{equation*}%
\begin{equation}
\dsum\limits_{n=0}^{\infty }\dsum\limits_{p=0}^{\infty }\frac{\Gamma \left(
n+m+1/2\right) }{\Gamma \left( p+n+m+3/2\right) \Gamma \left( n+2m+1\right) }%
\frac{1}{n!p!}\left( \frac{\gamma ^{2}k^{2}}{4}\right) ^{n}\left( -\frac{%
\gamma ^{2}}{4}\right) ^{p}  \label{EqnA20}
\end{equation}%
Expressing equation (\ref{EqnA20}) in terms of Pochhammer symbols gives
after some reduction the double hypergeometric series%
\begin{equation*}
\bar{\Lambda}_{-}^{m}\left( rR,\gamma ,k\right) =\frac{i}{\sqrt{rR}\left(
2m+1\right) !}\left( \frac{\gamma k}{2}\right) ^{2m+1}\times
\end{equation*}%
\begin{equation}
\dsum\limits_{n=0}^{\infty }\dsum\limits_{p=0}^{\infty }\frac{\left(
m+1/2\right) _{n}}{\left( m+3/2\right) _{n+p}\left( 2m+1\right) _{n}}\frac{1%
}{n!p!}\left( \frac{\gamma ^{2}k^{2}}{4}\right) ^{n}\left( \frac{-\gamma ^{2}%
}{4}\right) ^{p}  \label{EqnA21}
\end{equation}%
and this can be expressed as a Kamp\'{e} de F\'{e}riet function as defined
by Srivastava and Karlsson [12, p. 27]:%
\begin{equation*}
F_{l:m;n}^{p:q;r}\left[ 
\begin{array}{c}
\left\{ a_{p}\right\} : \\ 
\left\{ \alpha _{l}\right\} :%
\end{array}%
\begin{array}{c}
\left\{ b_{q}\right\} ; \\ 
\left\{ \beta _{m}\right\} ;%
\end{array}%
\begin{array}{c}
\left\{ c_{r}\right\} ; \\ 
\left\{ \gamma _{n}\right\} ;%
\end{array}%
x,y\right] =
\end{equation*}%
\begin{equation}
\dsum\limits_{j=0}^{\infty }\dsum\limits_{i=0}^{\infty }\frac{%
\dprod\limits_{u=1}^{p}\left( a_{u}\right)
_{j+i}\dprod\limits_{u=1}^{q}\left( b_{u}\right)
_{j}\dprod\limits_{u=1}^{r}\left( c_{u}\right) _{i}}{\dprod\limits_{u=1}^{l}%
\left( \alpha _{u}\right) _{j+i}\dprod\limits_{u=1}^{m}\left( \beta
_{u}\right) _{j}\dprod\limits_{u=1}^{n}\left( \gamma _{u}\right) _{i}}\frac{%
x^{j}}{j!}\frac{y^{i}}{i!}\text{.}  \label{EqnA22}
\end{equation}%
In the definition (\ref{EqnA22}), $\left\{ a_{p}\right\} \equiv
a_{1},..a_{p} $ and $\left\{ b_{q}\right\} $ and so on, are the lists of the
arguments of the Pochhammer symbols of the various types which appear in the
products on the right-hand side of the equation. If a list has no members,
it is represented by a hyphen. Comparing (\ref{EqnA21})\ with (\ref{EqnA22})
gives%
\begin{equation*}
\bar{\Lambda}_{-}^{m}\left( rR,\gamma ,k\right) =\frac{i}{\sqrt{rR}\left(
2m+1\right) !}\left( \frac{\gamma ^{2}k^{2}}{4}\right) ^{m+1/2}\times
\end{equation*}%
\begin{equation}
F_{1:1;0}^{0:1;0}\left[ 
\begin{array}{c}
-: \\ 
m+3/2:%
\end{array}%
\begin{array}{c}
m+1/2; \\ 
2m+1;%
\end{array}%
\begin{array}{c}
-; \\ 
-;%
\end{array}%
\frac{\gamma ^{2}k^{2}}{4},\frac{-\gamma ^{2}}{4}\right] \text{.}
\label{EqnA23}
\end{equation}

\section{Consequences of the hypergeometric formulas}

In the static limit as $\beta \rightarrow 0$ then equation (\ref{Eqn31})
reduces to%
\begin{equation}
\bar{\Lambda}_{+}^{m}\left( rR,0,k\right) =\frac{\Gamma \left( m+1/2\right) 
}{m!\sqrt{\pi rR}}\left( \frac{k}{2}\right) ^{2m+1}{}{}_{2}F_{1}\left(
m+1/2,m+1/2;2m+1;k^{2}\right)  \label{EqnB1}
\end{equation}%
and from (\ref{Eqn34}) and the standard hypergeometric identity [15, eqn
7.3.1 71]:%
\begin{equation}
{}_{2}F_{1}\left( a,b;2b;z\right) =\frac{2^{2b}}{\sqrt{\pi }}\frac{\Gamma
\left( b+1/2\right) }{\Gamma \left( 2b-a\right) }z^{-b}\left( 1-z\right)
^{\left( b-a\right) /2}\exp \left( i\pi \left( a-b\right) \right)
Q_{b-1}^{b-a}\left( \frac{2}{z}-1\right)  \label{EqnB2}
\end{equation}%
this reduces to 
\begin{equation}
\bar{\Lambda}_{+}^{m}\left( rR,0,k\right) =\frac{1}{\pi \sqrt{rR}}%
Q_{m-1/2}\left( \frac{2-k^{2}}{k^{2}}\right)  \label{EqnB3}
\end{equation}%
in agreement with equations (\ref{Eqn11})\ and (\ref{Eqn34}).

\subsection{Fourier coefficients as series of special functions}

The double series given by equation (\ref{Eqn31}) can be summed with respect
to either the index $n$ or the index $p$ in the definition (\ref{Eqn37}).
Summing with respect to $n$ in (\ref{Eqn37}) gives a series of Bessel
functions of the second kind:%
\begin{equation}
\bar{\Lambda}_{+}^{m}\left( rR,\gamma ,k\right) =\frac{-1}{2\sqrt{rR}}\left( 
\frac{k^{2}\gamma }{2}\right) ^{m+1/2}\dsum\limits_{p=0}^{\infty }\frac{%
\Gamma \left( m+1/2+p\right) }{\Gamma \left( 2m+1+p\right) p!}\left( \frac{%
k^{2}\gamma }{2}\right) ^{p}Y_{m+\frac{1}{2}+p}\left( \gamma \right)
\allowbreak  \label{EqnB4}
\end{equation}%
where the gamma function identity (\ref{Eqn46}) and the Bessel function
identity 
\begin{equation}
Y_{\nu }\left( \gamma \right) =\frac{1}{\sin \nu \pi }\left[ \cos \left( \nu
\pi \right) -J_{-\nu }\left( \gamma \right) \right]  \label{EqnB5}
\end{equation}%
have been employed to obtain equation (\ref{EqnB4}). Summing instead over
the index $p$ in (\ref{Eqn37}) gives the alternative series%
\begin{equation*}
\bar{\Lambda}_{+}^{m}\left( rR,\gamma ,k\right) =\frac{1}{m!\sqrt{\pi rR}}%
\left( \frac{k}{2}\right) ^{2m+1}\times
\end{equation*}%
\begin{equation}
\dsum\limits_{n=0}^{\infty }\Gamma \left( m-n+1/2\right) \frac{\left( \gamma
^{2}/4\right) ^{n}}{n!}\allowbreak {}{}_{2}F_{1}\left(
m+1/2-n,m+1/2;2m+1;k^{2}\right) \text{.}  \label{EqnB6}
\end{equation}%
This can be reduced using (\ref{EqnB2}) and (\ref{Eqn46}) to give:%
\begin{equation*}
\bar{\Lambda}_{+}^{m}\left( rR,\gamma ,k\right) =\frac{\left( -1\right) ^{m}%
}{\sqrt{rR}}\times
\end{equation*}%
\begin{equation}
\dsum\limits_{n=0}^{\infty }\left( \frac{\gamma ^{2}\sqrt{1-k^{2}}}{4}%
\right) ^{n}\frac{Q_{m-1/2}^{n}\left( \left( 2-k^{2}\right) /k^{2}\right) }{%
n!\Gamma \left( n-m+1/2\right) \Gamma \left( n+m+1/2\right) }\text{.}
\label{EqnB7}
\end{equation}%
The dimensionless variables $\gamma $ and $\lambda $ defined by equations (%
\ref{Eqn34}) and (\ref{Eqn40}) respectively are related by%
\begin{equation}
\gamma =\frac{\sqrt{2}\lambda }{k}  \label{EqnB8}
\end{equation}%
and substituting this equation and equation (\ref{Eqn36}) in equation (\ref%
{EqnB7}) gives immediately equation (\ref{Eqn44}).

Summing with respect to $p$ in equation (\ref{EqnA20}) gives the Bessel
series%
\begin{equation}
\bar{\Lambda}_{-}^{m}\left( rR,\gamma ,k\right) =\frac{i}{2\sqrt{rR}}%
\sum_{p=0}^{\infty }\frac{\Gamma \left( p+m+1/2\right) }{\Gamma \left(
p+2m+1\right) p!}\left( \frac{\gamma k^{2}}{2}\right) ^{p+m+1/2}J_{m+\frac{1%
}{2}+p}\left( \gamma \right)  \label{EqnB9}
\end{equation}%
$\allowbreak $The corresponding summation over the index $n$ gives%
\begin{equation*}
\bar{\Lambda}_{-}^{m}\left( rR,\gamma ,k\right) =\frac{i\sqrt{\pi }}{m!\sqrt{%
rR}}\left( \frac{\gamma k}{4}\right) ^{2m+1}\times
\end{equation*}%
\begin{equation}
\dsum\limits_{n=0}^{\infty }\frac{1}{\Gamma \left( n+m+3/2\right) n!}\left( 
\frac{-\gamma ^{2}}{4}\right) ^{n}{}{}_{1}F_{2}\left( m+1/2;n+m+3/2,2m+1;%
\frac{\gamma ^{2}k^{2}}{4}\right) \text{.}  \label{EqnB10}
\end{equation}%
There seems to be no hypergeometric transformation listed in standard tables
suitable for directly reducing the hypergeometric function in this equation.

Equations (\ref{EqnB4}) and (\ref{EqnB9}) can be conveniently combined to
give a series of Hankel functions of the first kind:%
\begin{equation}
\bar{G}_{H}^{m}\left( rR,\gamma ,k\right) =\frac{i}{2\sqrt{rR}}%
\dsum\limits_{p=0}^{\infty }\frac{\Gamma \left( p+m+1/2\right) }{\Gamma
\left( p+2m+1\right) p!}\left( \frac{\gamma k^{2}}{2}\right) ^{p+m+1/2}H_{m+%
\frac{1}{2}+p}^{\left( 1\right) }\left( \gamma \right) \text{,}
\label{EqnB11}
\end{equation}%
where $H_{m+\frac{1}{2}+p}^{\left( 1\right) }\left( \gamma \right) \equiv
J_{m+\frac{1}{2}+p}\left( \gamma \right) +iY_{m+\frac{1}{2}+p}\left( \gamma
\right) $.

\section{Differential equations from the two-dimensional hypergeometric
solutions}

Erd\'{e}lyi et. al. [11, section 5.9] tabulate the partial differential
equations satisfied by all the functions in Horn's list. They employ the
notation reproduced below for the various partial derivatives 
\begin{equation*}
p=\frac{\partial z}{\partial x};\text{ }q=\frac{\partial z}{\partial y};%
\text{ }r=\frac{\partial ^{2}z}{\partial x^{2}};\text{ }s=\frac{\partial
^{2}z}{\partial x\partial y};\text{ }t=\frac{\partial ^{2}z}{\partial y^{2}}
\end{equation*}%
where $z$ is any function on the list. Each function in the list satisfies
two partial differential equations, and unfortunately those given in [11,
5.9 34] for $z\left( \alpha ,\beta ,\delta ,x,y\right) \equiv \mathrm{H}%
_{3}\left( \alpha ,\beta ,\delta ,x,y\right) $ contain typographical errors.
The correct equations can be shown by the methods given in [11, section 5.7]
to be:%
\begin{equation}
x\left( 1-x\right) r+xys+\left[ \delta -\left( \alpha +\beta +1\right) x%
\right] p+\beta yq-\alpha \beta z=0  \label{EqnC0}
\end{equation}%
\begin{equation}
yt-xs+\left( 1-\alpha \right) q+z=0  \label{EqnC1}
\end{equation}%
and for the particular case considered here we have $\beta =\alpha $ and $%
\delta =2\alpha $ so that (\ref{EqnC0}) reduces to%
\begin{equation}
x\left( 1-x\right) r+xys+\left[ 2\alpha -\left( 2\alpha +1\right) x\right]
p+\alpha yq-\alpha ^{2}z=0\text{.}  \label{EqnC2}
\end{equation}%
Similar equations can also be derived for the Kamp\'{e} de F\'{e}riet
function defined by equation (\ref{Eqn38}). For the definition 
\begin{equation}
\bar{z}\left( \alpha ,u,v\right) \equiv F_{1:1;0}^{0:1;0}\left[ 
\begin{array}{c}
-: \\ 
\alpha +1:%
\end{array}%
\begin{array}{c}
\alpha ; \\ 
2\alpha ;%
\end{array}%
\begin{array}{c}
-; \\ 
-;%
\end{array}%
u,v\right]  \label{EqnC3}
\end{equation}%
and the notation%
\begin{equation*}
\bar{p}=\frac{\partial \bar{z}}{\partial u};\text{ }\bar{q}=\frac{\partial 
\bar{z}}{\partial v};\text{ }\bar{r}=\frac{\partial ^{2}\bar{z}}{\partial
u^{2}};\text{ }\bar{s}=\frac{\partial ^{2}\bar{z}}{\partial u\partial v};%
\text{ }\bar{t}=\frac{\partial ^{2}\bar{z}}{\partial v^{2}}
\end{equation*}%
then the equations corresponding to (\ref{EqnC1}) and (\ref{EqnC2}) can be
shown to be%
\begin{equation}
\bar{z}=v\bar{t}+u\bar{s}+\left( \alpha +1\right) \bar{q}  \label{EqnC4}
\end{equation}%
\begin{equation}
\bar{z}=2\alpha \bar{p}+u\bar{r}+\bar{q}+v\bar{t}.  \label{EqnC5}
\end{equation}

\subsection{Fourth-order differential equation from the Horn Function $%
\mathrm{H}_{3}\left( \protect\alpha ,\protect\beta ,\protect\delta %
,x,y\right) $}

Writing%
\begin{equation}
z\left( \alpha ,x,y\right) =\mathrm{H}_{3}\left( \alpha ,\alpha ,2\alpha
,x,y\right)  \label{EqnC6}
\end{equation}%
then (\ref{Eqn53}) can be expressed as%
\begin{equation}
z\left( \alpha ,x,y\right) =\frac{2^{2m+1}m!\sqrt{\pi }}{\Gamma \left(
m+1/2\right) }x^{-\alpha }g_{+}^{m}\left( x,y\right) \text{.}  \label{EqnC7}
\end{equation}%
The ordinary differential equation for $g_{+}^{m}\left( x,y\right) $ in
terms of $x\equiv k^{2}$ can be derived by first obtaining the corresponding
differential equation for $z\left( \alpha ,x,y\right) $ from (\ref{EqnC1})
and (\ref{EqnC2}) and then substituting (\ref{EqnC7}) into this equation.
Although straightforward in principle, this procedure is rather intricate in
practice, and only the essential elements of the derivation are given below.

\subsubsection{Differential equation for $z\left( \protect\alpha ,x,y\right) 
$}

It is convenient to define a differential operator $D$ such that:%
\begin{equation}
D\equiv x\frac{d}{dx}  \label{EqnC8}
\end{equation}%
which has by definition the properties:%
\begin{equation}
Dx=x  \label{EqnC9}
\end{equation}%
\begin{equation}
Dy=-y  \label{EqnC10}
\end{equation}%
\begin{equation}
D\left( xy\right) =0  \label{EqnC11}
\end{equation}%
\begin{equation}
Dz=xp-yq  \label{EqnC12}
\end{equation}%
\begin{equation}
Dp=xr-ys  \label{EqnC13}
\end{equation}%
\begin{equation}
Dq=xs-yt\text{.}  \label{EqnC14}
\end{equation}%
Combining equations (\ref{EqnC1}) and (\ref{EqnC14}) gives the equation 
\begin{equation}
Dq=z+\left( 1-\alpha \right) q  \label{EqnC19}
\end{equation}%
and applying the operators $\partial /\partial x$ and $\partial /\partial y$
to this equation gives respectively:%
\begin{equation}
Ds=p-\alpha s  \label{EqnC20}
\end{equation}%
\begin{equation}
Dt=q+\left( 2-\alpha \right) t\text{.}  \label{EqnC21}
\end{equation}%
Eliminating $\ $the variable $r$ gives a system of four coupled equations:

\begin{equation}
Dz+yq=xp  \label{EqnC22}
\end{equation}%
\begin{equation}
\left( 1-x\right) xDp=-xys+\left[ \left( 2\alpha +1\right) x-2\alpha \right]
xp-\alpha xyq+\alpha ^{2}xz  \label{EqnC23}
\end{equation}%
\begin{equation}
Dq=z+\left( 1-\alpha \right) q  \label{EqnC24}
\end{equation}%
\begin{equation}
xDs=xp-\alpha xs\text{.}  \label{EqnC25}
\end{equation}%
Eliminating $p$ gives a system of 3 equations:

\begin{equation}
D^{2}z-xD^{2}z+\left( 2\alpha -1\right) Dz-2\alpha xDz+yz-xyz-\alpha
^{2}xz=-xys+\left( 1-\alpha \right) yq  \label{EqnC26}
\end{equation}%
\begin{equation}
Dq=z+\left( 1-\alpha \right) q  \label{EqnC27}
\end{equation}%
\begin{equation}
xDs=Dz+yq-\alpha xs.  \label{EqnC28}
\end{equation}%
Eliminating $s$ gives the two equations:

\begin{equation*}
D^{3}z-xD^{3}z-\left( 3\alpha +1\right) xD^{2}z+\left( 3\alpha -1\right)
D^{2}z-\alpha \left( 3\alpha +2\right) xDz+2yDz
\end{equation*}%
\begin{equation}
+\alpha \left( 2\alpha -1\right) Dz-xyDz-\alpha ^{2}\left( \alpha +1\right)
xz+2\left( \alpha -1\right) yz-\alpha xyz=-y^{2}q  \label{EqnC29}
\end{equation}%
\begin{equation}
Dq=z+\left( 1-\alpha \right) q.  \label{EqnC30}
\end{equation}%
Eliminating $q$ gives finally the fourth-order equation:%
\begin{equation*}
\left( 1-x\right) D^{4}z+\left[ 4\alpha -\left( 4\alpha +3\right) x\right]
D^{3}z
\end{equation*}%
\begin{equation*}
+\left[ 5\alpha ^{2}+\alpha -1-\left( 6\alpha ^{2}+9\alpha +2\right) x-xy+2y%
\right] D^{2}z
\end{equation*}%
\begin{equation*}
+\left[ \alpha \left( 2\alpha -1\right) \left( \alpha +1\right) -\alpha
\left( 4\alpha ^{2}+9\alpha +4\right) x+2\left( 2\alpha -1\right) y-\left(
2\alpha +1\right) xy\right] Dz
\end{equation*}%
\begin{equation}
+\left[ 2\alpha \left( \alpha -1\right) y-\alpha ^{2}\left( \alpha +1\right)
\left( \alpha +2\right) x-\alpha \left( \alpha +1\right) xy+y^{2}\right] z=0%
\text{.}  \label{EqnC31}
\end{equation}%
This equation can be converted to standard differential form using the
identity:%
\begin{equation}
Dz\equiv x\frac{dz}{dx}  \label{EqnC32}
\end{equation}%
which gives:%
\begin{equation*}
\left( 1-x\right) x^{4}\frac{d^{4}z}{dx^{4}}+\left[ 2\left( 2\alpha
+3\right) -\left( 4\alpha +9\right) x\right] x^{3}\frac{d^{3}z}{dx^{3}}
\end{equation*}%
\begin{equation*}
+\left[ \left( \alpha +2\right) \left( 5\alpha +3\right) -3\left( \alpha
+2\right) \left( 2\alpha +3\right) x-xy+2y\right] x^{2}\frac{d^{2}z}{dx^{2}}
\end{equation*}%
\begin{equation*}
+\left[ 2\alpha \left( \alpha +1\right) \left( \alpha +2\right) -\left(
\alpha +1\right) \left( \alpha +2\right) \left( 4\alpha +3\right) x+4\alpha
y-2\left( \alpha +1\right) xy\right] x\frac{dz}{dx}
\end{equation*}%
\begin{equation}
+\left[ 2\alpha \left( \alpha -1\right) y-\alpha ^{2}\left( \alpha +1\right)
\left( \alpha +2\right) x-\alpha \left( \alpha +1\right) xy+y^{2}\right] z=0%
\text{.}  \label{EqnC33}
\end{equation}%
From equation (\ref{EqnC7}), the Fourier coefficient $g_{+}^{m}\left(
x,y\right) $ is related to $z\left( \alpha ,x,y\right) $ by

\begin{equation}
z=Cx^{-\alpha }g_{+}^{m}\left( x,y\right)  \label{EqnC34}
\end{equation}%
where the constant $C$ is given by equation (\ref{EqnC7}). Differentiating
equation (\ref{EqnC34}) gives the relations%
\begin{equation}
x\frac{dz}{dx}=Cx^{-\alpha }[-\alpha g_{+}^{m}+x\frac{dg_{+}^{m}}{dx}]
\label{EqnC35}
\end{equation}%
\begin{equation}
x^{2}\frac{d^{2}z}{dx^{2}}=Cx^{-\alpha }[\alpha \left( \alpha +1\right)
g_{+}^{m}-2\alpha x\frac{dg_{+}^{m}}{dx}+x^{2}\frac{d^{2}g_{+}^{m}}{dx^{2}}]
\label{EqnC36}
\end{equation}%
\begin{equation*}
x^{3}\frac{d^{3}z}{dx^{3}}=Cx^{-\alpha }\times
\end{equation*}%
\begin{equation}
\lbrack -\alpha \left( \alpha +1\right) \left( \alpha +2\right)
g_{+}^{m}+3\alpha \left( \alpha +1\right) x\frac{dg_{+}^{m}}{dx}-3\alpha
x^{2}\frac{d^{2}g_{+}^{m}}{dx^{2}}+x^{3}\frac{d^{3}g_{+}^{m}}{dx^{3}}]
\label{EqnC37}
\end{equation}%
\begin{equation*}
x^{4}\frac{d^{4}z}{dx^{4}}=Cx^{-\alpha }\times
\end{equation*}%
\begin{equation*}
\lbrack \alpha \left( \alpha +1\right) \left( \alpha +2\right) \left( \alpha
+3\right) g_{+}^{m}-4\alpha \left( \alpha +1\right) \left( \alpha +2\right) x%
\frac{dg_{+}^{m}}{dx}+
\end{equation*}%
\begin{equation}
6\alpha \left( \alpha +1\right) x^{2}\frac{d^{2}g_{+}^{m}}{dx^{2}}-4\alpha
x^{3}\frac{d^{3}g_{+}^{m}}{dx^{3}}+x^{4}\frac{d^{4}g_{+}^{m}}{dx^{4}}].
\label{EqnC38}
\end{equation}%
and substituting these relations into equation (\ref{Eqn33}) gives after
much reduction the fourth-order differential equation:%
\begin{equation*}
\left( 1-x\right) x^{4}\frac{d^{4}g_{+}^{m}}{dx^{4}}+\left( 6-9x\right) x^{3}%
\frac{d^{3}g_{+}^{m}}{dx^{3}}+\left[ 6-\alpha \left( \alpha -1\right)
-18x+y\left( 2-x\right) \right] x^{2}\frac{d^{2}g_{+}^{m}}{dx^{2}}+
\end{equation*}%
\begin{equation}
-2\left[ \alpha \left( \alpha -1\right) +x\left( y+3\right) \right] x\frac{%
dg_{+}^{m}}{dx}+y^{2}g_{+}^{m}=0\text{.}  \label{EqnC39}
\end{equation}

\subsection{Fourth-order differential equation from the Kamp\'{e} de F\'{e}%
riet function}

Equation (\ref{Eqn54}) can be written in the form%
\begin{equation}
g_{-}^{m}\left( x,y\right) =\frac{i}{\left( 2m+1\right) !}u^{\alpha }\bar{z}%
\left( \alpha ,u,v\right)  \label{EqnC40}
\end{equation}%
where 
\begin{equation}
u=xy  \label{EqnC41}
\end{equation}%
and 
\begin{equation}
v=-y\text{.}  \label{EqnC42}
\end{equation}%
The equation for $g_{-}^{m}\left( x,y\right) $ can be established in the
same manner as for equation (\ref{EqnC39}), but having already derived (\ref%
{EqnC39}), it is enough to establish that $g_{+}^{m}\left( x,y\right) $ and $%
g_{-}^{m}\left( x,y\right) $ both obey the same differential equation. From
equations (\ref{EqnC41})\ and (\ref{EqnC42}) then%
\begin{equation}
\frac{\partial }{\partial u}=\frac{1}{y}\frac{\partial }{\partial x}
\label{EqnC43}
\end{equation}%
\begin{equation}
\frac{\partial }{\partial v}=\frac{x}{y}\frac{\partial }{\partial x}-\frac{%
\partial }{\partial y}  \label{EqnC44}
\end{equation}%
and with the definition%
\begin{equation}
\bar{w}\left( \alpha ,x,y\right) =\bar{z}\left( \alpha ,xy,-y\right)
\label{EqnC45}
\end{equation}%
and the notation%
\begin{equation*}
\tilde{p}=\frac{\partial \bar{w}}{\partial x};\text{ }\tilde{q}=\frac{%
\partial \bar{w}}{\partial y};\text{ }\tilde{r}=\frac{\partial ^{2}\bar{w}}{%
\partial x^{2}};\text{ }\tilde{s}=\frac{\partial ^{2}\bar{w}}{\partial
x\partial y};\text{ }\tilde{t}=\frac{\partial ^{2}\bar{w}}{\partial y^{2}}
\end{equation*}%
then we have%
\begin{equation}
\bar{p}=\frac{\tilde{p}}{y}  \label{EqnC46}
\end{equation}%
\begin{equation}
\bar{q}=\frac{x}{y}\tilde{p}-\tilde{q}  \label{EqnC47}
\end{equation}%
\begin{equation}
\bar{r}=\frac{\tilde{r}}{y^{2}}  \label{EqnC48}
\end{equation}%
\begin{equation}
\bar{s}=\frac{x}{y^{2}}\tilde{r}+\frac{\tilde{p}}{y^{2}}-\frac{\tilde{s}}{y}
\label{EqnC49}
\end{equation}%
\begin{equation}
\bar{t}=\tilde{t}+2\frac{x}{y^{2}}\tilde{p}+\frac{x^{2}}{y^{2}}\tilde{r}-2%
\frac{x}{y}\tilde{s}\text{.}  \label{EqnC50}
\end{equation}%
Employing equations (\ref{EqnC45})-(\ref{EqnC50}) in the Kamp\'{e} de F\'{e}%
riet differential equations (\ref{EqnC4}) and (\ref{EqnC5}) gives 
\begin{equation}
y\bar{w}=-y^{2}\tilde{t}+xy\tilde{s}+\alpha x\tilde{p}-\left( \alpha
+1\right) y\tilde{q}  \label{EqnC51}
\end{equation}%
\begin{equation}
y\bar{w}=\left( 2\alpha -x\right) \tilde{p}+x\left( 1-x\right) \tilde{r}-y%
\tilde{q}-y^{2}\tilde{t}+2xy\tilde{s}\text{.}  \label{EqnC52}
\end{equation}%
To compare these equations with the partial differential equations from the
Horn function, we note that 
\begin{equation}
g_{+}^{m}\left( \alpha ,x,y\right) =Cx^{\alpha }z\left( \alpha ,x,y\right)
\label{EqnC53}
\end{equation}%
whereas%
\begin{equation}
g_{-}^{m}\left( \alpha ,x,y\right) =Ex^{\alpha }y^{\alpha }\bar{w}\left(
\alpha ,x,y\right)  \label{EqnC54}
\end{equation}%
where $C$ and $E$ are constants. Defining%
\begin{equation}
w\left( \alpha ,x,y\right) =y^{\alpha }\bar{w}\left( \alpha ,x,y\right)
\label{EqnC55}
\end{equation}%
then $g_{+}^{m}\left( x,y\right) $ and $g_{-}^{m}\left( x,y\right) $ will
satisfy the same ordinary differential equation if $z\left( \alpha
,x,y\right) $ and $w\left( \alpha ,x,y\right) $ satisfy the same pair of
partial differential equations. With the notation 
\begin{equation*}
\hat{p}=\frac{\partial w}{\partial x};\text{ }\hat{q}=\frac{\partial w}{%
\partial y};\text{ }\hat{r}=\frac{\partial ^{2}w}{\partial x^{2}};\text{ }%
\hat{s}=\frac{\partial ^{2}w}{\partial x\partial y};\text{ }\hat{t}=\frac{%
\partial ^{2}w}{\partial y^{2}}
\end{equation*}%
then%
\begin{equation}
\bar{w}\left( \alpha ,x,y\right) =y^{-\alpha }w\left( \alpha ,x,y\right)
\label{EqnC56}
\end{equation}%
\begin{equation}
\tilde{p}=y^{-\alpha }\hat{p}  \label{EqnC57}
\end{equation}%
\begin{equation}
\tilde{q}=y^{-\alpha }\left( \hat{q}-\frac{\alpha }{y}w\right)
\label{EqnC58}
\end{equation}%
\begin{equation}
\tilde{r}=y^{-\alpha }\hat{r}  \label{EqnC59}
\end{equation}%
\begin{equation}
\tilde{s}=y^{-\alpha }\left( \hat{s}-\frac{\alpha }{y}\hat{p}\right)
\label{EqnC60}
\end{equation}%
\begin{equation}
\tilde{t}=y^{-\alpha }\left( \hat{t}+\frac{\alpha \left( \alpha +1\right) }{%
y^{2}}w-2\frac{\alpha }{y}\hat{q}\right)  \label{EqnC61}
\end{equation}%
and substituting these relations in equations (\ref{EqnC51}) and (\ref%
{EqnC52}) gives%
\begin{equation}
y\hat{t}-x\hat{s}+\left( 1-\alpha \right) \hat{q}+w=0\text{.}  \label{EqnC62}
\end{equation}%
\begin{equation}
yw=\left[ 2\alpha -\left( 2\alpha +1\right) x\right] \hat{p}+x\left(
1-x\right) \hat{r}-y^{2}\hat{t}-\alpha ^{2}w+\left( 2\alpha -1\right) y\hat{q%
}+2xy\hat{s}\text{.}  \label{EqnC63}
\end{equation}%
Eliminating $yw$ from (\ref{EqnC63}) using equation (\ref{EqnC62}) gives%
\begin{equation}
x\left( 1-x\right) \hat{r}+xy\hat{s}+\left[ 2\alpha -\left( 2\alpha
+1\right) x\right] \hat{p}+\alpha y\hat{q}-\alpha ^{2}w=0\text{.}
\label{EqnC64}
\end{equation}%
Equations (\ref{EqnC62}) and (\ref{EqnC64}) obtained from the Kamp\'{e} de F%
\'{e}riet function are identical to equations (\ref{EqnC1}) and (\ref{EqnC2}%
) obtained from the Horn function, so $g_{-}^{m}\left( x,y\right) $ also
satisfies the differential equation (\ref{EqnC39}).

\section{Fourth-order differential equations from an integral representation}

The integral representation (\ref{Eqn9}) for the Fourier coefficient can be
written in the form:%
\begin{equation}
\hat{G}_{H}^{m}\left( rR,\lambda ,\omega \right) =\frac{\hat{y}_{m}\left(
\lambda ,\omega \right) }{\pi \sqrt{2rR}}  \label{EqnD1}
\end{equation}%
where%
\begin{equation}
\hat{y}_{m}\left( \lambda ,\omega \right) =\dint\limits_{0}^{\pi }\frac{\exp
\left( i\lambda \sqrt{\omega -\cos \psi }\right) }{\sqrt{\omega -\cos \psi }}%
\cos \left( m\psi \right) d\psi \text{.}  \label{EqnD2}
\end{equation}%
The function $y_{m}\left( \lambda ,\omega \right) $ satisfies the partial
differential equation%
\begin{equation}
\frac{\partial ^{2}y_{m}\left( \lambda ,\omega \right) }{\partial \omega
\partial \lambda }=-\frac{\lambda }{2}y_{m}\left( \lambda ,\omega \right)
\label{EqnD3}
\end{equation}%
and this has the elementary separated solution:%
\begin{equation}
Y_{m}\left( s,\lambda ,\omega \right) =C_{m}\left( s\right) \exp \left[ \pm
i\left( s\omega +\frac{\lambda ^{2}}{4s}\right) \right]  \label{EqnD4}
\end{equation}%
where $s$ is the separation constant. The solution $y_{m}\left( \lambda
,\omega \right) $ can be constructed as a superposition of the allowable
(i.e. finite at infinity) elementary solutions given by (\ref{EqnD4}). This
gives $y_{m}\left( \lambda ,\omega \right) $ in the form:%
\begin{equation}
y_{m}(\lambda ,\omega )=\dint\limits_{0}^{\infty }C_{m}(s)\exp \left[ \pm
i\left( s\omega +\frac{\lambda ^{2}}{4s}\right) \right] ds  \label{EqnD5}
\end{equation}%
where the $\pm $ sign is chosen so that the integral converges. As $\omega $
is real and positive, this depends only on the imaginary part of $\lambda
\equiv \alpha +i\sigma $, the appropriate sign being the same as that of $%
\sigma $, which will be assumed positive here. Setting $\lambda =0$ in (\ref%
{EqnD2})\ and (\ref{EqnD5}) gives 
\begin{equation}
\dint\limits_{0}^{\infty }C_{m}(s)\exp \left( is\omega \right) ds=\sqrt{2}%
Q_{m-1/2}(\omega )  \label{EqnD6}
\end{equation}%
and $C_{m}(s)$ can be determined from the integral [7, eqn 6.621 1]:%
\begin{equation*}
\dint\limits_{0}^{\infty }\exp \left( -s\delta \right) J_{\nu }\left(
s\right) s^{\mu -1}ds=
\end{equation*}%
\begin{equation}
\frac{\Gamma \left( \nu +\mu \right) }{2^{\nu }\delta ^{\mu +\nu }\Gamma
\left( \nu +1\right) }{}_{2}F_{1}\left( \frac{\nu +\mu }{2},\frac{\nu +\mu +1%
}{2};\nu +1;-\frac{1}{\delta ^{2}}\right) \text{.}  \label{EqnD7}
\end{equation}%
Setting $\delta =-i\omega $, $\nu =m$ and $\mu =1/2$ in equation (\ref{EqnD7}%
) gives%
\begin{equation*}
\dint\limits_{0}^{\infty }\exp \left( i\omega s\right) J_{m}\left( s\right)
s^{-1/2}ds=
\end{equation*}%
\begin{equation}
\frac{\Gamma \left( m+1/2\right) }{\left( -i\right) ^{m+1/2}2^{m}\omega
^{m+1/2}\Gamma \left( m+1\right) }{}{}_{2}F_{1}\left( \frac{m+1/2}{2},\frac{%
m+3/2}{2};m+1;\frac{1}{\omega ^{2}}\right)  \label{EqnD8}
\end{equation}%
and the expression for $Q_{m-1/2}\left( \omega \right) $ in terms of the
Gauss hypergeometric function \ is [16, eqn 8.1.3]%
\begin{equation}
Q_{m-1/2}\left( \omega \right) =\frac{1}{2^{m}}\sqrt{\frac{\pi }{2}}\frac{%
\Gamma \left( m+1/2\right) }{\omega ^{m+1/2}\Gamma \left( m+1\right) }{}{}%
_{2}F_{1}\left( \frac{m+1/2}{2},\frac{m+3/2}{2};m+1;\frac{1}{\omega ^{2}}%
\right) \text{.}  \label{EqnD9}
\end{equation}%
From equations (\ref{EqnD6}), (\ref{EqnD8})\ and (\ref{EqnD9}) it follows
that%
\begin{equation}
C_{m}(s)=\sqrt{\pi }\left( -i\right) ^{m+1/2}J_{m}\left( s\right) s^{-1/2}
\label{EqnD10}
\end{equation}%
and hence%
\begin{equation}
y_{m}(\lambda ,\omega )=\sqrt{\pi }\left( -i\right)
^{m+1/2}\dint\limits_{0}^{\infty }\exp \left[ i\left( \omega s+\frac{\lambda
^{2}}{4s}\right) \right] J_{m}\left( s\right) s^{-1/2}ds\text{.}
\label{EqnD11}
\end{equation}%
For the special case of evanescent waves such that $\lambda =i\sigma $ with $%
\sigma >0$ then with a suitable transformation in the complex plane, this
equation can be expressed in the form%
\begin{equation}
y_{m}(i\sigma ,\omega )=\sqrt{\pi }\dint\limits_{0}^{\infty }\exp \left[
-\left( \omega s+\frac{\sigma ^{2}}{4s}\right) \right] I_{m}\left( s\right)
s^{-1/2}ds.  \label{EqnD12}
\end{equation}%
The details of this transformation are given in Appendix B. Equation (\ref%
{EqnD12}) is straightforward to evaluate numerically as the integrand is not
oscillatory and decays exponentially to zero as $s\rightarrow \infty $
provided $\omega >1$, which from equation (\ref{Eqn12}) is always the case.
This follows immediately from the leading term in the asymptotic
approximation as $s\rightarrow \infty $ of $I_{m}\left( s\right) $, which is
[7, eqn 8.451 5]:%
\begin{equation}
I_{m}\left( s\right) \sim \frac{\exp \left( s\right) }{\sqrt{2\pi s}}\text{.}
\label{EqnD12a}
\end{equation}

\subsection{Fourth-order differential equation in terms of $\protect\omega $}

The integral representation (\ref{EqnD11}) allows the ordinary differential
equations in terms of $\omega $ or $\lambda $ satisfied by $y_{m}(\lambda
,\omega )$ to be constructed in a straightforward manner. It is convenient
to define a new variable $\chi $ such that 
\begin{equation}
\chi =\frac{\lambda ^{2}}{4}  \label{EqnD13}
\end{equation}%
and also a new dependent variable $\hat{y}_{m}(\chi ,\omega )$ such that 
\begin{equation}
\hat{y}_{m}(\chi ,\omega )\equiv y_{m}\left( \lambda ,\omega \right) \text{.}
\label{EqnD14}
\end{equation}%
Then $\hat{y}_{m}(\chi ,\omega )$ is given by 
\begin{equation}
\hat{y}_{m}(\chi ,\omega )=\dint\limits_{0}^{\infty }J_{m}(s)f(s,\omega
,\chi )ds  \label{EqnD15}
\end{equation}%
where $\chi $ is to be regarded as a constant embedded parameter in the ODE\
satisfied by $\hat{y}_{m}(\chi ,\omega ),$ and where $f(s,\omega ,\chi )$ is
given by%
\begin{equation}
f(s,\omega ,\chi )\equiv \sqrt{\pi }\left( -i\right) ^{m+1/2}s^{1/2}\exp %
\left[ i\left( \omega s+\frac{\chi }{s}\right) \right] \text{.}
\label{EqnD16}
\end{equation}%
The various derivatives of $\hat{y}\left( \omega ,\chi \right) $ are then
given by%
\begin{equation}
\frac{d^{n}\hat{y}_{m}}{d\omega ^{n}}\equiv \left( i\right)
^{n}\dint\limits_{0}^{\infty }s^{n}J_{m}(s)f(s,\omega ,\chi )ds\text{.}
\label{EqnD17}
\end{equation}%
The Bessel function $J_{m}(s)$ satisfies the differential equation [7, eqn
8.401]%
\begin{equation}
\frac{1}{s}\frac{d}{ds}\left( s\frac{dJ_{m}\left( s\right) }{ds}\right)
=\left( \frac{m^{2}}{s^{2}}-1\right) J_{m}\left( s\right)  \label{EqnD18}
\end{equation}%
and therefore:%
\begin{equation}
\dint\limits_{0}^{\infty }\left( \frac{m^{2}}{s^{2}}-1\right) J_{m}\left(
s\right) f(s,\omega ,\chi )ds=\dint\limits_{0}^{\infty }\frac{1}{s}\frac{d}{%
ds}\left( s\frac{dJ_{m}\left( s\right) }{ds}\right) f(s,\omega ,\chi )ds.
\label{EqnD19}
\end{equation}%
Differentiating (\ref{EqnD19}) twice with respect to $\omega $ and utilizing
equations (\ref{EqnD15})-(\ref{EqnD17})\ gives:%
\begin{equation*}
\dint\limits_{0}^{\infty }\left( m^{2}-s^{2}\right) J_{m}\left( s\right)
f(s,\omega ,\chi )ds=\dint\limits_{0}^{\infty }s\frac{d}{ds}\left( s\frac{%
dJ_{m}\left( s\right) }{ds}\right) f(s,\omega ,\chi )ds
\end{equation*}%
\begin{equation}
\frac{d^{2}\hat{y}_{m}}{d\omega ^{2}}+m^{2}\hat{y}_{m}=\dint\limits_{0}^{%
\infty }s\frac{d}{ds}\left( s\frac{dJ_{m}\left( s\right) }{ds}\right)
f(s,\omega ,\chi )ds.  \label{EqnD20}
\end{equation}%
Integrating this twice by parts yields%
\begin{equation}
\frac{d^{2}\hat{y}_{m}}{d\omega ^{2}}+m^{2}\hat{y}_{m}=\dint\limits_{0}^{%
\infty }J_{m}(s)\frac{d}{ds}\left( s\frac{d}{ds}\left[ sf(s,\omega ,\chi )%
\right] \right) ds.  \label{EqnD21}
\end{equation}%
Since%
\begin{equation}
\frac{d}{ds}\left( s\frac{d}{ds}\left[ sf(s,\omega ,\chi )\right] \right)
=\left( \frac{1}{4}+2i\omega s-\omega ^{2}s^{2}+2\omega \chi -\chi
^{2}s^{-2}\right) f(s,\omega ,\chi )  \label{EqnD22}
\end{equation}%
then%
\begin{equation}
\frac{d^{2}\hat{y}_{m}}{d\omega ^{2}}+m^{2}\hat{y}_{m}=\dint\limits_{0}^{%
\infty }J_{m}(s)\left( \frac{1}{4}+2i\omega s-\omega ^{2}s^{2}+2\omega \chi
-\chi ^{2}s^{-2}\right) f(s,\omega ,\chi )ds.  \label{EqnD23}
\end{equation}%
For the Poisson case with $\chi =0$, employing (\ref{EqnD17}) in (\ref%
{EqnD23}) gives Legendre's equation (\ref{Eqn57}) of degree $m-1/2$, as must
be the case for consistency with equation (\ref{Eqn11}). For the Helmholtz
case, (\ref{EqnD23})\ must be differentiated twice with respect to $\omega $
before employing (\ref{EqnD17}). This yields the fourth-order linear ODE%
\begin{equation}
(1-\omega ^{2})\frac{d^{4}\hat{y}_{m}}{d\omega ^{4}}-6\omega \frac{d^{3}\hat{%
y}_{m}}{d\omega ^{3}}+(m^{2}-2\chi \omega -\frac{25}{4})\frac{d^{2}\hat{y}%
_{m}}{d\omega ^{2}}-4\chi \frac{d\hat{y}_{m}}{d\omega }-\chi ^{2}\hat{y}%
_{m}=0  \label{EqnD25}
\end{equation}%
which becomes equation (\ref{Eqn55}) on substituting $\chi \equiv \lambda
^{2}/4$.

\subsection{Equivalence of the differential equations}

Equation (\ref{EqnD25}) can be converted to an equation in terms of $x\equiv
k^{2}$ by making the substitutions:%
\begin{equation}
\chi =\frac{xy}{2}  \label{EqnD26}
\end{equation}%
\begin{equation}
\omega =\frac{2-x}{x}  \label{EqnD27}
\end{equation}%
\begin{equation}
1-\omega ^{2}=\frac{4\left( x-1\right) }{x^{2}}  \label{EqnD28}
\end{equation}%
\begin{equation}
m^{2}-2\chi \omega -\frac{25}{4}=\alpha ^{2}-\alpha -6+\left( x-2\right) y
\label{EqnD29}
\end{equation}%
\begin{equation}
\frac{dy}{d\omega }=-\frac{x^{2}}{2}\frac{d\bar{y}}{dx}  \label{EqnD30}
\end{equation}%
\begin{equation}
\frac{d^{2}y}{d\omega ^{2}}=\frac{x^{4}}{4}\frac{d^{2}\bar{y}}{dx^{2}}+\frac{%
x^{3}}{2}\frac{d\bar{y}}{dx}  \label{EqnD31}
\end{equation}%
\begin{equation}
\frac{d^{3}\bar{y}}{d\omega ^{3}}=-\frac{x^{6}}{8}\frac{d^{3}\bar{y}}{dx^{3}}%
-\frac{3x^{5}}{4}\frac{d^{2}\bar{y}}{dx^{2}}-\frac{3x^{4}}{4}\frac{d\bar{y}}{%
dx}  \label{EqnD32}
\end{equation}%
\begin{equation}
\frac{d^{4}y}{d\omega ^{4}}=\frac{x^{8}}{16}\frac{d^{4}\bar{y}}{dx^{4}}+%
\frac{3x^{7}}{4}\frac{d^{3}\bar{y}}{dx^{3}}+\frac{9x^{6}}{4}\frac{d^{2}\bar{y%
}}{dx^{2}}+\frac{3x^{5}}{2}\frac{d\bar{y}}{dx}\text{.}  \label{EqnD33}
\end{equation}%
On collecting terms and simplifying this yields

\begin{equation*}
\left( 1-x\right) x^{4}\frac{d^{4}\bar{y}}{dx^{4}}+\left( 6-9x\right) x^{3}%
\frac{d^{3}\bar{y}}{dx^{3}}+
\end{equation*}%
\begin{equation*}
\left[ \alpha -\alpha ^{2}+6-18x+\left( 2-x\right) y\right] x^{2}\frac{d^{2}%
\bar{y}}{dx^{2}}+
\end{equation*}%
\begin{equation}
-2\left[ \alpha \left( \alpha -1\right) +x\left( 3+y\right) \right] x\frac{d%
\bar{y}}{dx}+y^{2}\bar{y}=0\text{.}  \label{EqnD34}
\end{equation}%
Inspection of equations (\ref{EqnC39}) and (\ref{EqnD34}), obtained by
totally different methods, shows that they are identical.

\section{Comments and conclusions}

The Fourier coefficients for the Helmholtz Green function have been split
into their half advanced+half retarded and half advanced$-$half retarded
components, and these components have been given in closed form in terms of
two-dimensional hypergeometric functions. These solutions generalize the
well-known solutions of Poisson's equation for ring sources, and reduce to
them in the static limit when the wave number $\beta =0$. The
two-dimensional hypergeometric functions can be considered as double series,
with the order of summation arbitrary. The two summation choices give
different series of special functions for each of the Fourier components,
and all of these series have been numerically verified, as have the closed
form solutions themselves. One series is given in terms of Hankel functions,
and only a few terms are need far from the ring source for accurate results.
A second series in terms of associated Legendre functions only requires a
few terms in the neigborhood of the ring to give accurate results.

The systems of partial differential equations associated with each of the
two generalized hypergeometric functions have been used to derive a
fourth-order ordinary differential equation in terms of $x\equiv k^{2}$ for
the Fourier coefficients. A completely different approach involving integral
representations of the Fourier coefficients has been presented in tandem,
which derives many of the same results, as well as some new ones. Both
approaches give exactly the same fourth-order differential equation for the
general Fourier coefficient, despite the algebra being rather intricate in
both cases. Another fourth order ordinary differential equation in terms of
the wave number parameter $\lambda $ can also be derived by the methods
presented here.

\appendix

\section{Series from an integral representation}

The Fourier coefficient $G_{H}^{m}\left( \beta ,r,R,z-Z\right) \equiv $ $%
\hat{G}_{H}^{m}\left( rR,\lambda ,\omega \right) $ given by equations (\ref%
{Eqn24}) and (\ref{Eqn25}) can be expressed in the form%
\begin{equation}
\hat{G}_{H}^{m}\left( rR,\lambda ,\omega \right) =\frac{1}{\pi \sqrt{2rR}}%
\dsum\limits_{n=0}^{\infty }\dint\limits_{0}^{\pi }\frac{\left( i\lambda
\right) ^{n}}{n!}\left( \omega -\cos \psi \right) ^{\left( n-1\right)
/2}\cos \left( m\psi \right) d\psi  \label{App01}
\end{equation}%
where $\omega $ and $\lambda $ are defined by equations (\ref{Eqn12}) and (%
\ref{Eqn40}) respectively. Evaluation of (\ref{App01}) requires the integral%
\begin{equation}
\hat{I}_{m,n}\left( \omega \right) =\dint\limits_{0}^{\pi }\cos \left( m\psi
\right) \left( \omega -\cos \psi \right) ^{\left( n-1\right) /2}d\psi
\label{App02}
\end{equation}%
which can be evaluated for $m\in 
\mathbb{N}
_{0}$ using the integral representation [7, eqn 8.711 2]:%
\begin{equation}
P_{\nu }^{m}\left( \xi \right) =\frac{\Gamma \left( \nu +m+1\right) }{\pi
\Gamma \left( \nu +1\right) }\dint\limits_{0}^{\pi }\cos \left( m\theta
\right) \left( \xi +\sqrt{\xi ^{2}-1}\cos \theta \right) ^{\nu }d\theta
\label{App03}
\end{equation}%
which is equivalent to 
\begin{equation}
P_{\nu }^{m}\left( \xi \right) =\frac{\left( -1\right) ^{m}\Gamma \left( \nu
+m+1\right) \left( \xi ^{2}-1\right) ^{\nu /2}}{\pi \Gamma \left( \nu
+1\right) }\dint\limits_{0}^{\pi }\cos \left( m\psi \right) \left( \frac{\xi 
}{\sqrt{\xi ^{2}-1}}-\cos \psi \right) ^{\nu }d\psi \text{.}  \label{App04}
\end{equation}%
The substitutions%
\begin{equation}
\omega =\frac{\xi }{\sqrt{\xi ^{2}-1}}  \label{App05}
\end{equation}%
and%
\begin{equation}
\nu =\frac{n-1}{2}  \label{App06}
\end{equation}%
then give%
\begin{equation}
\hat{I}_{m,n}\left( \omega \right) =\frac{\left( -1\right) ^{m}\pi \Gamma
\left( \left( n+1\right) /2\right) \left( \omega ^{2}-1\right) ^{\left(
n-1\right) /4}}{\Gamma \left( m+\left( n+1\right) /2\right) }P_{\left(
n-1\right) /2}^{m}\left( \frac{\omega }{\sqrt{\omega ^{2}-1}}\right)
\label{App07}
\end{equation}%
and (\ref{App01}) becomes%
\begin{equation*}
\hat{G}_{H}^{m}\left( rR,\lambda ,\omega \right) =\frac{\left( -1\right) ^{m}%
}{\left( \omega ^{2}-1\right) ^{1/4}\sqrt{2rR}}\times
\end{equation*}%
\begin{equation}
\dsum\limits_{n=0}^{\infty }\frac{\left( i\lambda \left( \omega
^{2}-1\right) ^{1/4}\right) ^{n}}{n!}\frac{\Gamma \left( \left( n+1\right)
/2\right) }{\Gamma \left( m+\left( n+1\right) /2\right) }P_{\left(
n-1\right) /2}^{m}\left( \frac{\omega }{\sqrt{\omega ^{2}-1}}\right) \text{.}
\label{App08}
\end{equation}%
Splitting the series (\ref{App08}) into even and odd terms gives after some
reduction%
\begin{equation*}
\hat{\Lambda}_{+}^{m}\left( rR,\lambda ,\omega \right) =\frac{\left(
-1\right) ^{m}\sqrt{\pi }}{\left( \omega ^{2}-1\right) ^{1/4}\sqrt{2rR}}%
\times
\end{equation*}%
\begin{equation}
\dsum\limits_{p=0}^{\infty }\left( \frac{-\lambda ^{2}\sqrt{\omega ^{2}-1}}{4%
}\right) ^{p}\frac{1}{p!\Gamma \left( p+m+1/2\right) }P_{p-1/2}^{m}\left( 
\frac{\omega }{\sqrt{\omega ^{2}-1}}\right)  \label{App09}
\end{equation}%
\begin{equation*}
\hat{\Lambda}_{-}^{m}\left( rR,\lambda ,\omega \right) =\frac{\left(
-1\right) ^{m}\sqrt{\pi }i\lambda }{2\sqrt{2rR}}\times
\end{equation*}%
\begin{equation}
\dsum\limits_{s=m}^{\infty }\left( \frac{-\lambda ^{2}\sqrt{\omega ^{2}-1}}{4%
}\right) ^{s}\frac{1}{\Gamma \left( s+3/2\right) \Gamma \left( m+s+1\right) }%
P_{s}^{m}\left( \frac{\omega }{\sqrt{\omega ^{2}-1}}\right) \text{,}
\label{App10}
\end{equation}%
where the factorial formulas 
\begin{equation}
\left( 2p\right) !=\frac{2^{2p}\Gamma \left( p+1/2\right) p!}{\sqrt{\pi }}
\label{App11}
\end{equation}%
\begin{equation}
\left( 2s+1\right) !=\frac{2^{2s+1}\Gamma \left( s+3/2\right) s!}{\sqrt{\pi }%
}  \label{App12}
\end{equation}%
have been used to obtain (\ref{App09}) and (\ref{App10}). The index in (\ref%
{App10}) runs from $s=m$ rather than from $s=0$ as the associated Legendre
polynomial $P_{s}^{m}\left( \xi \right) $ is zero for $s<m$. The
substitution $s=p+m$ in (\ref{App10}) gives the alternative form%
\begin{equation*}
\hat{\Lambda}_{-}^{m}\left( rR,\lambda ,\omega \right) =\frac{\sqrt{\pi }%
i\lambda }{2\sqrt{2rR}}\left( \frac{\lambda ^{2}\sqrt{\omega ^{2}-1}}{4}%
\right) ^{m}\times
\end{equation*}%
\begin{equation}
\dsum\limits_{p=0}^{\infty }\left( \frac{-\lambda ^{2}\sqrt{\omega ^{2}-1}}{4%
}\right) ^{p}\frac{1}{\Gamma \left( p+m+3/2\right) \Gamma \left(
p+2m+1\right) }P_{p+m}^{m}\left( \frac{\omega }{\sqrt{\omega ^{2}-1}}\right) 
\text{.}  \label{App13}
\end{equation}%
The indices in equations (\ref{App09})\ and (\ref{App13}) can be switched to
negative values using the relations [16, eqns 8.2.5, 8.2.1]:%
\begin{equation}
P_{\nu }^{m}\left( \xi \right) =\frac{\Gamma \left( \nu +m+1\right) }{\Gamma
\left( \nu -m+1\right) }P_{\nu }^{-m}\left( \xi \right) \text{ }\left[ m\in 
\mathbb{N}
_{0}\right]  \label{App14}
\end{equation}%
\begin{equation}
P_{-\nu -1}^{m}\left( \xi \right) =P_{\nu }^{m}\left( \xi \right)
\label{App15}
\end{equation}%
which gives%
\begin{equation*}
\hat{\Lambda}_{+}^{m}\left( rR,\lambda ,\omega \right) =\frac{\left(
-1\right) ^{m}\sqrt{\pi }}{\left( \omega ^{2}-1\right) ^{1/4}\sqrt{2rR}}%
\times
\end{equation*}%
\begin{equation}
\dsum\limits_{p=0}^{\infty }\left( \frac{-\lambda ^{2}\sqrt{\omega ^{2}-1}}{4%
}\right) ^{p}\frac{1}{p!\Gamma \left( p-m+1/2\right) }P_{-p-1/2}^{-m}\left( 
\frac{\omega }{\sqrt{\omega ^{2}-1}}\right)  \label{App16}
\end{equation}%
\begin{equation*}
\hat{\Lambda}_{-}^{m}\left( rR,\lambda ,\omega \right) =\frac{\sqrt{\pi }%
i\lambda }{2\sqrt{2rR}}\left( \frac{\lambda ^{2}\sqrt{\omega ^{2}-1}}{4}%
\right) ^{m}\times
\end{equation*}%
\begin{equation}
\dsum\limits_{p=0}^{\infty }\left( \frac{-\lambda ^{2}\sqrt{\omega ^{2}-1}}{4%
}\right) ^{p}\frac{1}{\Gamma \left( p+m+3/2\right) \Gamma \left( p+1\right) }%
P_{-p-m-1}^{-m}\left( \frac{\omega }{\sqrt{\omega ^{2}-1}}\right) \text{.}
\label{App17}
\end{equation}%
The kind of Legendre functions in equations (\ref{App16}) and (\ref{App17})
can be switched using the Whipple relation [16, eqn 8.2.7],[17]:%
\begin{equation}
P_{-\mu -1/2}^{-\alpha -1/2}\left( \frac{\omega }{\sqrt{\omega ^{2}-1}}%
\right) =\frac{\left( \omega ^{2}-1\right) ^{1/4}\exp \left( -i\mu \pi
\right) }{\left( \pi /2\right) ^{1/2}\Gamma \left( \alpha +\mu +1\right) }%
Q_{\alpha }^{\mu }\left( \omega \right)  \label{App18}
\end{equation}%
which gives after some reduction%
\begin{equation*}
\hat{\Lambda}_{+}^{m}\left( rR,\lambda ,\omega \right) =\frac{\left(
-1\right) ^{m}}{\sqrt{rR}}\times
\end{equation*}%
\begin{equation}
\dsum\limits_{p=0}^{\infty }\left( \frac{\lambda ^{2}\sqrt{\omega ^{2}-1}}{4}%
\right) ^{p}\frac{Q_{m-1/2}^{p}\left( \omega \right) }{p!\Gamma \left(
p-m+1/2\right) \Gamma \left( m+p+1/2\right) }  \label{App19}
\end{equation}%
\begin{equation*}
\hat{\Lambda}_{-}^{m}\left( rR,\lambda ,\omega \right) =\frac{\left(
-1\right) ^{m}}{\sqrt{rR}}\times
\end{equation*}%
\begin{equation}
\dsum\limits_{p=0}^{\infty }\left( \frac{\lambda ^{2}\sqrt{\omega ^{2}-1}}{4}%
\right) ^{p+m+1/2}\frac{Q_{m-1/2}^{p+m+1/2}\left( \omega \right) }{p!\Gamma
\left( p+m+3/2\right) \Gamma \left( p+2m+1\right) }\text{.}  \label{App20}
\end{equation}

\section{Transformation of an integral in the complex plane}

The integral for $y_{m}\left( i\sigma ,\omega \right) $ in equation (\ref%
{EqnD11}) for $\lambda =i\sigma $ can be considered to be the contribution
along the real axis of the contour integral%
\begin{equation}
T_{m}\left( \sigma ,\omega \right) =\sqrt{\pi }\left( -i\right)
^{m+1/2}\doint\limits_{C}\exp \left[ i\left( \omega s-\frac{\sigma ^{2}}{4s}%
\right) \right] J_{m}\left( s\right) s^{-1/2}ds  \label{AppB1}
\end{equation}%
where $C$ is the closed contour shown in the figure below, in the limits as $%
\epsilon \rightarrow 0$ and $R\rightarrow \infty $.

\FRAME{fhFU}{3.25in}{3.0831in}{0pt}{\Qcb{Contour C in the complex $s-$plane
where $s=x+iy$. The smaller quarter circle is of radius $\protect\epsilon %
\rightarrow 0$ and the larger quarter circle is of radius $R\rightarrow
\infty $. The integrand of the contour integral has an isolated essential
singularity at the origin of the $s-$plane and is analytic on and within the
contour $C$.}}{\Qlb{Figure1}}{figure1.ps}{\special{language "Scientific
Word";type "GRAPHIC";maintain-aspect-ratio TRUE;display "USEDEF";valid_file
"F";width 3.25in;height 3.0831in;depth 0pt;original-width
7.7574in;original-height 11.0627in;cropleft "0.2460";croptop "1";cropright
"0.9384";cropbottom "0.5396";filename
'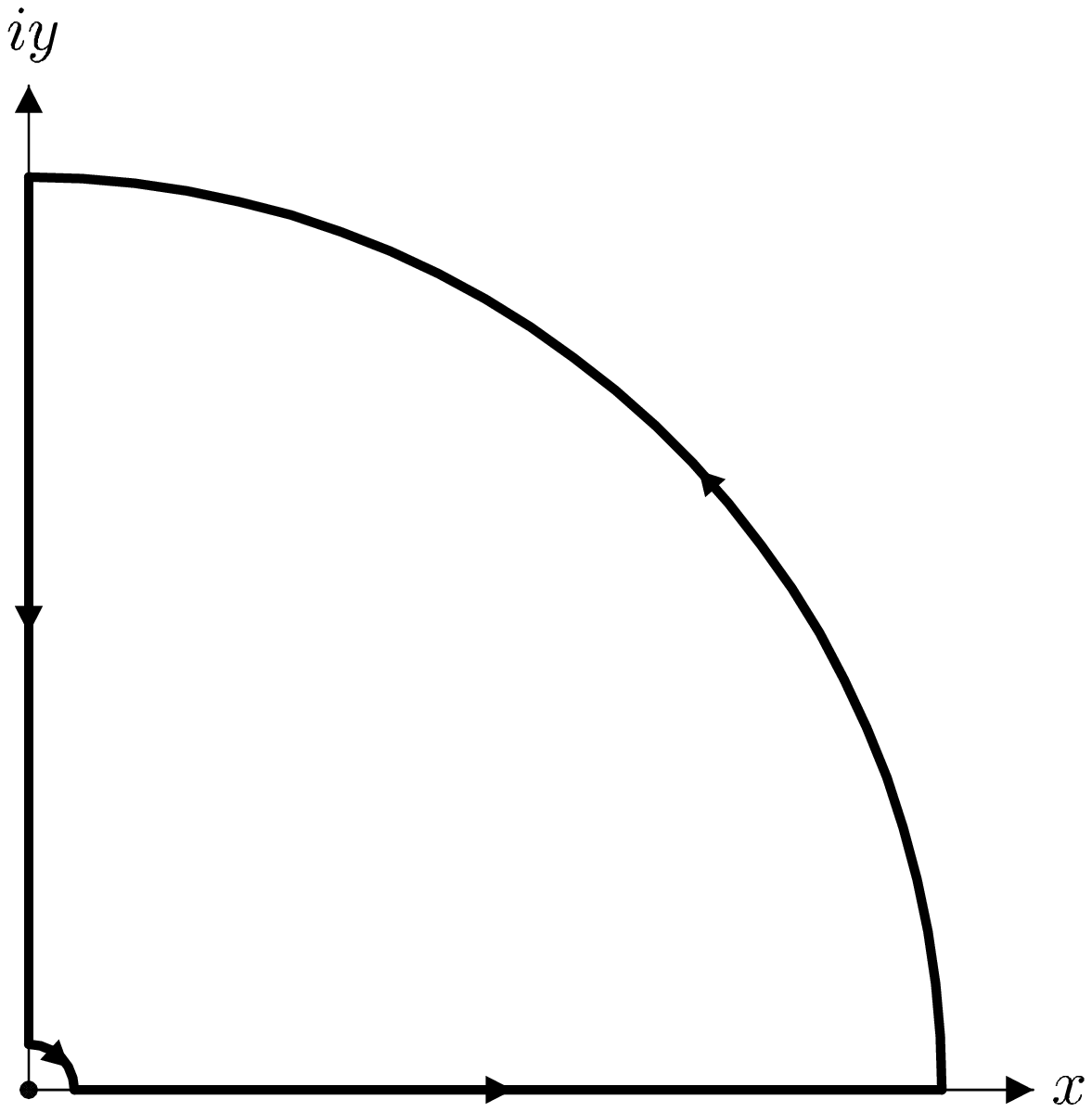';file-properties "XNPEU";}}The corresponding
contribution to the contour integral along the imaginary axis is given by%
\begin{equation}
U_{m}\left( \omega ,\sigma \right) =-\sqrt{\pi }\left( -i\right)
^{m+1/2}\dint\limits_{0}^{\infty }\exp \left[ -\left( \omega y+\frac{\sigma
^{2}}{4y}\right) \right] J_{m}\left( iy\right) y^{-1/2}\left( i\right)
^{1/2}dy  \label{AppB2}
\end{equation}%
and this can be stated in terms of the modified Bessel function of the first
kind using the identity [7, eqn 8.406 3]%
\begin{equation}
J_{m}\left( iy\right) =\left( i\right) ^{m}I_{m}\left( y\right)
\label{AppB3}
\end{equation}%
which gives immediately%
\begin{equation}
U_{m}\left( \sigma ,\omega \right) =-\sqrt{\pi }\dint\limits_{0}^{\infty
}\exp \left[ -\left( \omega y+\frac{\sigma ^{2}}{4y}\right) \right]
I_{m}\left( y\right) y^{-1/2}dy\text{.}  \label{AppB4}
\end{equation}%
The integrand of (\ref{AppB1}) is analytic everywhere within the contour $C$
and therefore from Cauchy's theorem, the integral (\ref{AppB1}) is zero.
Therefore if the contributions to the contour integral along the two quarter
circles vanish in the limits as $\epsilon \rightarrow 0$ and $R\rightarrow
\infty $, then%
\begin{equation}
y_{m}\left( i\sigma ,\omega \right) +U_{m}\left( \sigma ,\omega \right) =0
\label{AppB5}
\end{equation}%
and equation (\ref{EqnD12}) has been proven. Along the smaller quarter
circle we set $s=\epsilon \exp \left( i\theta \right) $ and the contribution
to the contour integral becomes%
\begin{equation*}
V_{m}\left( \sigma ,\omega \right) =-\sqrt{\pi }\left( -i\right)
^{m+1/2}\dint\limits_{0}^{\pi /2}\exp \left[ i\left( \omega \epsilon \exp
\left( i\theta \right) -\frac{\sigma ^{2}}{4\epsilon }\exp \left( -i\theta
\right) \right) \right] \times
\end{equation*}%
\begin{equation}
J_{m}\left( \epsilon \exp \left( i\theta \right) \right) i\epsilon
^{1/2}d\theta  \label{AppB6}
\end{equation}%
and this clearly vanishes as $\epsilon \rightarrow 0$. Along the larger
quarter circle we set $s=R\exp \left( i\theta \right) $ and the contribution
to the integral is given by%
\begin{equation*}
W_{m}\left( \sigma ,\omega \right) =\sqrt{\pi }\left( -i\right)
^{m+1/2}\dint\limits_{0}^{\pi /2}\exp \left[ i\left( \omega R\exp \left(
i\theta \right) -\frac{\sigma ^{2}}{4R}\exp \left( -i\theta \right) \right) %
\right] \times
\end{equation*}%
\begin{equation}
J_{m}\left( R\exp \left( i\theta \right) \right) iR^{1/2}\exp \left( i\theta
/2\right) d\theta \text{.}  \label{AppB7}
\end{equation}%
The leading term in the asymptotic approximation of $J_{m}\left( s\right) $
is given by [7, eqn 8.451 1]%
\begin{equation}
J_{m}\left( s\right) =\sqrt{\frac{2}{\pi s}}\cos \left( s-\frac{\pi m}{2}-%
\frac{\pi }{4}\right)  \label{AppB8}
\end{equation}%
and therefore%
\begin{equation*}
W_{m}\left( \sigma ,\omega \right) =\sqrt{2}\left( -i\right)
^{m+1/2}\dint\limits_{0}^{\pi /2}\exp \left[ i\left( \omega R\exp \left(
i\theta \right) -\frac{\sigma ^{2}}{4R}\exp \left( -i\theta \right) \right) %
\right] \times
\end{equation*}%
\begin{equation}
\frac{\left( i\right) ^{1/2}}{2}\left[ \exp \left( iR\exp \left( i\theta
\right) -\frac{i\pi m}{2}-\frac{i\pi }{4}\right) +\exp \left( -iR\exp \left(
i\theta \right) +\frac{i\pi m}{2}+\frac{i\pi }{4}\right) \right] d\theta
\label{AppB9}
\end{equation}%
which can be expressed as%
\begin{equation*}
\frac{\left( -i\right) ^{m}}{\sqrt{2}}\dint\limits_{0}^{\pi /2}\left[ \exp
\left( -i\frac{\sigma ^{2}}{4R}\exp \left( -i\theta \right) \right) \right]
\times
\end{equation*}%
\begin{equation}
\left( \exp \left[ i\left( \left( \omega +1\right) R\exp \left( i\theta
\right) -\frac{i\pi m}{2}-\frac{i\pi }{4}\right) \right] +\exp \left[
i\left( \left( \omega -1\right) R\exp \left( i\theta \right) +\frac{i\pi m}{2%
}+\frac{i\pi }{4}\right) \right] \right) d\theta \text{.}  \label{AppB10}
\end{equation}%
Inspection of (\ref{AppB10}) shows that the integrand vanishes exponentially
in the limit as $R\rightarrow \infty $ provided $\omega >1$. This condition
always holds and is also the condition for the integral in (\ref{EqnD12}) to
converge.

\section{Numerical results}

The series solutions for the Fourier coefficients given by equation (\ref%
{Eqn51}) and equations (\ref{Eqn44})-(\ref{Eqn45}) were evaluated using
Mathematica and the numerical performance was explored for various geometric
parameters and wave numbers. For comparison, the two integrals (\ref{Eqn9})
and (\ref{Eqn19}) for the Fourier coefficients were also evaluated
numerically for the same parameters. All four methods give identical results
at locations which are neither too far away nor too close to the ring
source. The numerical integration (\ref{Eqn9}) performs very well at all
distances from the ring, whereas the numerical integration (\ref{Eqn19})
fails when either very close to the ring or too far away. No cases were
identified where equation (\ref{Eqn19}) was superior. The Hankel function
series (\ref{Eqn51}) requires fewer and fewer terms for convergence as the
distance from the loop increases, and conversely performance decreases as
the ring is approached. The associated Legendre function series (\ref{Eqn44}%
) and (\ref{Eqn45}) have precisely the opposite performance, with great
accuracy close to the ring and failure at large distances from the ring. The
two series (\ref{Eqn44}) and (\ref{Eqn45}) are well suited to calculations
close to the ring as the associated Legendre functions themselves each
contain the ring singularity as $\omega \rightarrow 1$. By contrast, the
Hankel functions (\ref{Eqn51}) are not singular at the ring and hence an
increasing number of terms are required to model the singularity as the ring
is approached. In all cases, there is are always at least one numerical
integration and one series solution which can be used to cross check each
other.

Sample numerical results are given in Table 2 for moderate distances from
the ring source, and shows the number of terms required by each series to
match the numerical integrations exactly. Table 3 shows the performance of
the Hankel series with increasing distance from the ring. The number of
Hankel terms decreases to very few at large distances from the ring. Table 4
shows the performance of the two associated Legendre series (\ref{Eqn44})
and (\ref{Eqn45}) as the ring is approached. The real part of $%
G_{H}^{m}\left( \beta ,r,R,z-Z\right) $ diverges logarithmically as $%
z\rightarrow Z$, whereas the imaginary part tends to a finite limit. It can
be seen immediately from Table 4 that only 8 terms in each Legendre series
is sufficient to calculate $G_{H}^{m}\left( \beta ,r,R,z-Z\right) $ for the
range $\{0<z\leqslant 1\}$.

\begin{table}[tbp]
$%
\begin{bmatrix}
m & \beta & r & z & N_{1} & N_{2} & G_{H}^{m}\left( \beta ,r,R,z-Z\right) \\ 
0 & 2 & 1/2 & 1/2 & 94 & 10 & -0.4332208795+0.6063507453\text{ }i \\ 
0 & 2 & 1/2 & 3/2 & 24 & 12 & -0.4324244083-0.2593676946\text{ }i \\ 
0 & 2 & 1/2 & 5 & 9 & 24 & -0.1320141290-0.1413052175\text{ }i \\ 
0 & 2 & 1/2 & 10 & 8 & 37 & 0.02892833221+0.09482283693\text{ }i \\ 
0 & 2 & 1/2 & 20 & 6 & 63 & -0.03552779935+0.03502697525\text{ }i \\ 
3 & 5 & 3/2 & 1/2 & 224 & 24 & -0.2152817201-0.2085849956\text{ }i \\ 
3 & 5 & 3/2 & 1 & 97 & 25 & 0.1382226177-0.2010980843\text{ }i \\ 
3 & 5 & 3/2 & 5 & 17 & 47 & -0.009794158906-0.000546039281\text{ }i \\ 
3 & 5 & 3/2 & 10 & 13 & 80 & -0.0004846328044+0.0006340532520\text{ }i \\ 
3 & 5 & 3/2 & 20 & 10 & 147 & 0.00000356468968+0.00005347913049\text{ }i%
\end{bmatrix}%
$%
\caption{Numerical solution for the Fourier coefficient $G_{H}^{m}\left( 
\protect\beta ,r,R,z-Z\right) $. $R=1$, $Z=0$ and the other parameters are
given in the table. $N_{1}$ is the number of terms in the Hankel function
series (\protect\ref{Eqn51}) needed to give the accuracy given. $N_{2}$ is
the number of terms in the associated Legendre series (\protect\ref{Eqn44})
and (\protect\ref{Eqn45}) to provide the accuracy given. Of the two
associated Legendre series, (\protect\ref{Eqn44}) requires more terms than (%
\protect\ref{Eqn45}) for the stated accuracy.}
\end{table}

\begin{table}[tbp]
$%
\begin{bmatrix}
z & N_{1} & G_{H}^{m}\left( \beta ,r,R,z-Z\right) \\ 
1/2 & 227 & 0.0785417676-0.2281496125\text{ }i \\ 
1 & 99 & 0.1318397799+0.0959755332\text{ }i \\ 
5 & 20 & 0.04717552085-0.09819984770\text{ }i \\ 
50 & 7 & -0.001762722093-0.000313668126\text{ }i \\ 
100 & 6 & -0.0000246835014+0.0004487208692\text{ }i \\ 
200 & 5 & -4.36384289\times 10^{-6}-0.00011237773355\text{ }i \\ 
1000 & 4 & -1.884281670\times 10^{-6}-4.086433833\times 10^{-6}\text{ }i \\ 
5000 & 3 & -1.446923432\times 10^{-7}+1.070704963\times 10^{-7}\text{ }i \\ 
10000 & 2 & 4.307310056\times 10^{-8}+1.302718185\times 10^{-8}\text{ }i \\ 
10^{7} & 1 & -2.303180610\times 10^{-14}+3.865922798\times 10^{-14}\text{ }i%
\end{bmatrix}%
$%
\caption{Series solution for the Fourier coefficient $G_{H}^{m}\left( 
\protect\beta ,r,R,z-Z\right) $. $m=1,$ $\protect\beta =6$, $R=1$, $Z=0$ $\ $%
and $r=3/2$. $z$ is given in the table. $N_{1}$ is the number of terms in
the Hankel function series (\protect\ref{Eqn51}) needed for the accuracy
given. }
\end{table}

\begin{table}[tbp]
$%
\begin{bmatrix}
z & N_{2} & G_{H}^{m}\left( \beta ,r,R,z-Z\right) \\ 
1 & 8 & 0.1874175169+0.1222388714\text{ }i \\ 
10^{-1} & 8 & 0.8955546890+0.1360159497\text{ }i \\ 
10^{-2} & 8 & 1.628566013+0.136158894\text{ }i \\ 
10^{-3} & 7 & 2.361506874+0.136160324\text{ }i \\ 
10^{-4} & 7 & 3.094442571+0.136160339\text{ }i \\ 
10^{-5} & 7 & 3.827378171+0.136160339\text{ }i \\ 
10^{-6} & 7 & 4.560313770+0.136160339\text{ }i \\ 
10^{-7} & 7 & 5.293249369+0.136160339\text{ }i \\ 
10^{-8} & 7 & 6.026184968+0.136160339\text{ }i \\ 
10^{-9} & 7 & 6.759120567+0.136160339\text{ }i%
\end{bmatrix}%
$%
\caption{Series solution for the Fourier coefficient $G_{H}^{m}\left( 
\protect\beta ,r,R,z-Z\right) $. $m=2,$ $\protect\beta =1$, $R=1$, $Z=0$ $\ $%
and $r=1$. $z$ is given in the table, and as $z\rightarrow 0$, the ring is
approached. $N_{2}$ is the number of terms in the associated Legendre
function series (\protect\ref{Eqn44}) and (\protect\ref{Eqn45}) needed for
the accuracy given. }
\end{table}


\begin{thebibliography}{99}
\bibitem{Ref001} J. Mathews, and R. L. Walker, \textit{Mathematical Methods
of Physics}, 2nd Edn, Addison Wesley, New York 1973.

\bibitem{Ref002} P. L. Overfelt, Near fields of the constant current thin
circular loop antenna of arbitrary radius, \textit{IEEE Trans. Antennas and
Propagat.}, \textbf{44} (1996) 166-171.

\bibitem{Ref003} D. H. Werner, An exact integration procedure for vector
potentials of\ thin circular loop antennas,\ \textit{IEEE\ Trans. Antennas
Propagat.}, \textbf{44} (1996) 157-165.

\bibitem{Ref004} J. T. Conway, New exact solution procedure for the near
fields of the thin\ circular loop antenna, \textit{IEEE\ Trans. Antennas
Propagat.}, \textbf{53} (2005) 509-517.

\bibitem{Ref005} P. R. Prentice, The acoustic ring source and its
application to propeller acoustics, \textit{Proc. R. Soc. Lond. A}, \textbf{%
437} (1992) 629-644.

\bibitem{Ref006} G. Matviyenko, On the azimuthal Fourier components of the
Green's function for the Helmholtz equation in three dimensions, \textit{J.
Math. Phys.}, \textbf{36} (1995) 5159-5169.

\bibitem{Ref007} I. S. Gradshteyn and I. M. Rhyzik,\textit{\ Table of
Integrals, Series and Products}, 7th Edn, Academic, New York 2007.

\bibitem{Ref008} H. S. Cohl, and J. E. Tohline, A compact cylindrical
Green's function expansion for the solution of potential problems, \textit{%
Astrophys. J.}, \textbf{527} (1999) 86-101.

\bibitem{Ref009} G. N. Watson, \textit{A Treatise on the Theory of Bessel
Functions}, 2nd Edn, Cambridge 1944.

\bibitem{Ref010} P. M. Morse and H. Feshbach, \textit{Methods of Theoretical
Physics vol. 1},\ McGraw-Hill, New York 1953.

\bibitem{Ref011} A. Erd\'{e}lyi, et al., \textit{Higher Transcendental
Functions vol. 1}, McGraw-Hill, New York 1953.

\bibitem{Ref012} H. M. Srivastava and P. W. Karlsson, \textit{Multiple
Gaussian Hypergeometric Series}, Ellis Horwood, Chichester 1985.

\bibitem{Ref013} J. T. Conway, Fourier series for elliptic integrals and
some generalizations via hypergeometric series, \textit{Intgr. Transf. Spec.
F.}, \textbf{19} (2008) 305-315.

\bibitem{Ref014} S. Wolfram, \textit{The Mathematica Book}, 5th Edn,Wolfram
Media, Champaign, IL, 2003.

\bibitem{Ref015} A. P. Prudnikov, Yu. A. Brychkov and O. I. Marichev, 
\textit{Integrals and Series\ vol. 3, More Special Functions}, Gordon and
Breach, New York 1990.

\bibitem{Ref016} M. Abramowitz and I. S. Stegun, \textit{Handbook of
Mathematical Functions},\ Dover, New York 1972.

\bibitem{Ref017} H. S. Cohl, J. E. Tohline, A. R. P. Rau and H. M.
Srivastava, Developments in determining the gravitational potential using
toroidal functions, \textit{Astrom. Nachr.} \textbf{321} (2000) 5/6, 363-372.
\end{thebibliography}
\end{document}